\documentclass[fleqn,usenatbib]{mnras}
\usepackage{graphicx}
\usepackage{amsmath,amssymb}
\usepackage{float}
\usepackage{multirow}
\usepackage{widetext}
\graphicspath{ {JLA_figures/} }
\def\etal{{et al.}}\def\LA{\Lambda}\def\LCDM{$\LA$CDM}\def\ts{time\-scape}
\def\Z#1{_{\lower1pt\hbox{$\scriptstyle#1$}}} \def\w#1{\,\hbox{#1}}
\def\X#1{_{\lower1pt\hbox{$\scriptscriptstyle#1$}}}
\def\Ns#1{_{\lower1pt\hbox{$\scriptstyle\rm#1$}}}
\def\lcdm{\hbox{$\scriptstyle\LA$}CDM}
\def\lsim{\mathop{\hbox{${\lower3.8pt\hbox{$<$}}\atop{\raise0.2pt\hbox{$\sim$}}
$}}} \def\gsim{\mathop{\hbox{${\lower3.8pt\hbox{$>$}}\atop{\raise0.2pt\hbox{$
\sim$}}$}}} \def\goesas{\mathop{\sim}\limits} \def\DE{\Delta}
\def\kms{\w{km}\;\w{sec}^{-1}}\def\kmsMpc{\kms\w{Mpc}^{-1}}
\def\h{\,h^{-1}}\def\hm{\h\hbox{Mpc}}\def\etBg{\et\Z{B\ga}}
\def\art#1{\ #1.\ }\def\ns#1{_{\rm #1}}
\def\AaA#1{A\&A,\ {\bo#1}} \def\AJ#1{AJ, {\bo#1}} \def\ApJ#1{ApJ,\ {\bo#1}}
\def\AJ#1{AJ, {\bo#1}} \def\MNRAS#1{MNRAS,\ {\bo#1}}
\def\PRL#1{Phys.\ Rev.\ Lett.,\ {\bo#1}} \def\PR#1#2{Phys.\ Rev.\ #1,\ {\bo#2}}
\def\ApJs#1{ApJS,\ {\bo#1}}
\def\CQG#1{Class.\ Quantum Grav.,\ {\bo#1}}
\def\GRG#1{Gen.\ Relativ.\ Grav.,\ {\bo#1}}
\def\JCAP#1{J.\ Cosmol.\ Astropart.\ Phys., {#1}, }
\def\bo{\bf} 
\def\art#1{} 
\def\etBg{\eta\Z{B\gamma}}\def\si{\sigma}\def\th{\theta}\def\al{\alpha}
\def\OmMn{\Omega\Z{M0}}\def\OmLn{\Omega\Z{\LA 0}}\def\be{\beta}
\def\Omkn{\Omega\Z{k0}}\def\OmRn{\Omega\Z{R0}}\def\CC{{\tt c}}
\def\OmM{\Omega\Z{M}}\def\OmL{\Omega\Z{\LA}}\def\dd{{\rm d}}
\def\Omk{\Omega\Z{k}}\def\OmR{\Omega\Z{R}}\def\Hn{H\Z0}
\def\dL{d\Z{L}}\def\dA{d\Z{A}}\def\dH{d\Z{H}}
\def\MB{M\Ns B}\def\MBn{M\Ns{B,0}}\def\siMB{\si\Z{M_{\rm B,0}}}
\def\mb{m\Ns B^*}\def\hmb{{\hat m}\Ns B^*}\def\fv{{f\ns v}}
\def\fvn{f\ns{v0}} \def\fvf{\left(1-\fv\right)}
\def\mean#1{{\vphantom{\tilde#1}\bar#1}} 
\def\bH{\mean H}\def\Hb{\bH\Z{\!0}}\def\zmin{z\ns{min}}
\def\ab{\mean a}\def\gb{\mean\gamma}
\def\frn#1#2{{\textstyle{#1\over#2}}} \def\tb{t'}
\def\ave#1{\langle#1\rangle}\def\pt{\partial}
\def\mulcdm{\mu\Z{\Lambda{\rm CDM}}}\def\FF{{\cal F}}
\def\zh{\hat z}\def\la{\lambda}\def\Mmod{\mathrm{M}}
\def\beq{\begin{equation}}\def\eeq{\end{equation}}
\def\hblank#1{\hbox to#1 mm{\hfil}}
\begin{document}
\catcode`@=11 \def\tod@y{}\def\@pubyear{2017}
\gdef\@journal{{\bf https://doi.org/10.1093/mnras/stx1858} \hfill MNRAS (2017)}
\def\@volume{\hbox{\hskip-2pt} DOI: 10.1093/mnras/stx1858}\def\@pagerange{}\catcode`@=12
\def\today{Accepted 2017 July 19. Received 2017 July 18; in original form 2017 June 22}
\title[Apparent cosmic acceleration from supernovae Ia]{Apparent cosmic acceleration from type Ia supernovae}
\author[Dam, Heinesen \& Wiltshire]{Lawrence H.~Dam, Asta Heinesen
and David L. Wiltshire\thanks{E-mail:
david.wiltshire@canterbury.ac.nz}\\
Department of Physics \& Astronomy, University of Canterbury,
Private Bag 4800, Christchurch 8140, New Zealand\\
}

\maketitle
\label{firstpage}

\begin{abstract}
Parameters that quantify the acceleration of cosmic expansion are conventionally
determined within the standard Friedmann-Lema\^{\i}tre-Robertson-Walker (FLRW)
model, which fixes spatial curvature to be homogeneous. Generic averages of
Einstein's equations in
inhomogeneous cosmology lead to models with non-rigidly evolving average spatial
curvature, and different parametrizations of apparent cosmic acceleration.
The \ts\ cosmology is a viable example of such a model without dark energy.
Using the largest available supernova data set, the JLA catalogue, we find
that the \ts\ model fits the luminosity distance-redshift data with a
likelihood that is statistically indistinguishable from the standard spatially
flat $\LA$ cold dark matter cosmology by Bayesian comparison. In the \ts\ case
cosmic acceleration is non-zero but has a marginal amplitude, with best-fitting
apparent deceleration parameter, $q\Z0=-0.043^{+0.004}_{-0.000}$.
Systematic issues regarding standardization of supernova light curves are
analysed. Cuts of data at the statistical homogeneity scale affect light
curve parameter fits independent of cosmology. A cosmological
model dependence of empirical changes to the mean colour parameter is also
found. Irrespective of which model ultimately fits better,
we argue that as a competitive model with a non-FLRW expansion history, the
\ts\ model may prove a useful diagnostic tool for disentangling selection
effects and astrophysical systematics from the underlying expansion history.
\end{abstract}
\begin{keywords}
cosmological parameters ---dark energy --- cosmology: observations ---
cosmology: theory
\end{keywords}

\section{Inhomogeneous alternatives to dark energy}

One of the most important discoveries in cosmology \citep{R98,P98} is the
observation that the luminosity distances
and redshifts of type Ia supernovae (SneIa) are well
matched to the expansion history of a spatial homogeneous and isotropic
Friedmann-Lema\^{\i}tre-Robertson-Walker (FLRW) model only if the
Universe began an epoch of accelerated expansion late in cosmic history. Since
gravity with matter obeying the Strong Energy Condition is universally
attractive, this
demands a cosmological constant, $\LA$, or some other unknown source of
spatially homogeneous dark energy with an equation of state, $P=w\rho c^2$,
that subsequent tests find to be close to the cosmological constant case,
$w=-1$. Independently of SneIa, since the late 1990s other data sets
including the cosmic microwave background (CMB) and galaxy clustering
statistics, have been found to independently require late epoch cosmic
acceleration in the FLRW model.

Despite this success, the nature of dark energy remains a mystery for
fundamental physics.
Furthermore, a number of puzzles remain \citep{lcdm1,lcdm2}
which range from significant anomalies \citep{lithium} to lesser tensions
\citep{LymanA}. While some puzzles may inevitably boil down to statistical
sampling biases, systematic problems in data reduction or potentially unknown
astrophysics, they may also in part be due to incorrect cosmological model
assumptions. It is therefore important to consider alternatives to the
standard cosmology, particularly any alternatives which do not change known
gravitational physics but which reconsider assumptions that are not demanded
theoretically in light of observations.
The assumption of spatial homogeneity and isotropy falls in this category.
While the remarkable isotropy of the CMB points to an initial state with
a very high degree of smoothness,
the late epoch Universe encompasses a complex cosmic web of structures
\citep{web}. It is dominated in volume by voids that are threaded and
surrounded by clusters of galaxies distributed in sheets, knots and
filaments.

Spatial homogeneity is first encountered in some statistical sense in
averages on scales $\gsim70$--$120\hm$ based on the 2-point galaxy correlation
function \citep{h05,sdb12}. This indicates that we can expect
cosmology to be modelled by some averaged description on larger scales,
but does not guarantee that such a description is given at all times
by the FLRW geometry, however. Demanding FLRW geometry on arbitrarily large
scales in space and time presupposes one
particular answer to the fitting problem \citep{fit1,fit2,dust}. The
fitting problem is an unanswered fundamental question in general
relativity: on what scale(s) are matter and geometry dynamically
coupled via Einstein's equations? Since general relativity is a causal
theory, the speed of light yields a natural upper limit. However, most of the
energy density of the present epoch Universe in the energy-momentum tensor
on the right hand side of Einstein's equations is in non-relativistic
matter subject to speeds much smaller than that of light. Consequently, the
effective causal scale for density fluctuations is the {\em matter
horizon} \citep{es09}, the comoving region which has significantly contributed
matter to the local physical environment of a given observer.

The matter horizon gives a natural effective causal upper limit for
direct applicability of the Einstein equations. Regardless of what the scale
is, if Einstein's equations apply only on small scales then generic large-scale
averages do not lead to the Einstein
equations for a prescribed averaged matter source. Instead one obtains averaged
cosmological models with backreaction \citep{be1,be2,BR}, whose evolution
generically
can differ significantly from that of the FLRW geometry. Since the cosmic
expansion history is phenomenologically very well described by a FLRW model
with a small number of parameters, then any physically relevant model of
backreaction must be subject to some simplifying principles that restrict
many possible mathematical choices for defining the average expansion history
and its relation to local observables \citep{bc1,bc2}.

The \ts\ scenario \citep*{clocks,sol,obs,dnw} provides such principles. Within
Buchert's scalar averaging formalism \citep{be1,be2}, the relationship between
statistical volume averages and local observables is restricted by
applying the Cosmological Equivalence Principle \citep{cep}, a generalization
of the Strong Equivalence Principle.
The clocks and rulers which best describe average cosmic evolution (the
`bare' or `volume--average' quantities) in Buchert's statistical
formalism are reinterpreted, and differ systematically from local
clocks and rulers of observers in environments where the matter density
differs significantly from the volume average. Given the smooth initial
state of the Universe, this only becomes significant late in cosmic history,
when spatial curvature gradients grow between bound structures (which
by definition are greater than critical density) and the voids which
dominate by volume.

The \ts\ model has passed a number of independent observational tests
\citep*{lnw,sw,grb,dnw,nw}. Its distance--redshift relation is very close to
that of particular $\LA$ cold dark matter (\LCDM) models over small redshift
ranges, but effectively
interpolates \citep{obs,bscg} between spatially flat \LCDM\ cosmologies with
different values of $\OmMn$ and $\OmLn$ over larger redshift ranges. In
particular, when the timescape model is fit to the angular diameter distance
of the sound horizon in the CMB then the spatially flat \LCDM\ model with the
closest comoving distance at redshifts $z\gsim100$ has a value
of $\OmMn$ 15--27\% lower than that of the spatially flat \LCDM\ model with
the closest comoving distance at redshifts $z<1.2$ \citep{obs,bscg}.

Geometric tests of the timescape expansion history are most developed
\citep{obs}, and give rise to measures \citep*{cbl,rbf} which will
definitely distinguish both the \ts\ model and other alternatives
\citep*{l09,lrs} from the standard FLRW model using
{\em Euclid} satellite data \citep*{smn}. On the other hand, tests of the
CMB anisotropy spectrum in the \ts\ model are at present limited by systematic uncertainties
of 8--13\% in parameters which relate to the matter content \citep{nw}.
This is a consequence of backreaction schemes having not yet been applied to
the primordial plasma.

\section{Supernova redshift-distance analysis}

In the case of the redshift range probed by SneIa the difference between
the \ts\ and \LCDM\ cosmologies is comparable to the systematic
uncertainties that arise between different methods for fitting the light
curves of SneIa to obtain ``standard candles''. In particular, in the last
full analysis of the \ts\ model (using data available in 2010) \citet{sw}
found significant differences between data reduced by the
MLCS2k2 (Multicolor Light-Curve Shape) fitter \citep{JRK}
and the SALT/SALT2 (Spectral
Adaptive Light-curve Template) fitters \citep{Guy05,Guy07}. While the
relative Bayesian evidence was
sometimes `positive' (but not very strong), the conclusion as to which
cosmological model fitted better depended on the light-curve fitting method.
Consequently the empirical nature of light-curve fitting may mask effects
due to the underlying expansion history if this deviates from
the FLRW geometry.

The significantly larger
Joint Light-curve Analysis (JLA) SneIa catalogue \citep{jla} now
makes possible a renewed comparison of the \ts\ and \LCDM\ models, as well
as further investigation of the systematics of light-curve fitting.
Recently, \citet*{ngs} (NGS16) have used the JLA catalogue
to reinvestigate systematic issues associated with SneIa light-curve fitting
within FLRW cosmologies using the SALT2 method. They adopted
maximum likelihood estimators (MLE) that take
into account the underlying Gaussianity of particular light-curve
parameters \citep{mtbsv}.
NGS16 concluded that the significance for cosmic
acceleration, as compared to an empty Milne model (or any cosmology with
constant expansion), is ``marginal'' (at $\lsim3\si$ significance). This
conclusion was challenged by \citet*{rh} (RH16), who introduced
12 additional light-curve parameters to allow for possible unaccounted
systematics, concluding that the $2.8\si$ significance found by NGS16
increased to $3.7\si$ for a general FLRW model, or to
$4.2\si$ for the spatially flat case. However, RH16 did not consider whether
the increased model complexity was justified from a Bayesian standpoint.

In the SALT2 method each
observed supernova redshift is used to determine a theoretical
distance modulus,
\beq \label{eq:mu}
\mu \equiv 25 + 5 \log_{10}\left( \dL \over \w{Mpc}\right)\,,
\eeq
where $\dL$ is the luminosity
distance for each cosmological model. This is then
compared to the observed distance modulus, which is related to the
supernova light-curve by
\beq
\mu\Ns{SN}=\mb-\MB+\al x\Z1-\be c,
\label{salt}\eeq
where $m\Ns B^*$ is the apparent magnitude at maximum in the rest-frame $B$ band,
$\MB$ is the corresponding absolute magnitude of the source, $x_1$ and $c$
are empirical parameters that describe the light-curve stretch and colour
corrections for each supernova, while $\al$ and $\be$ are parameters that are
assumed to be constant for {\em all} SneIa.

The theoretical distance modulus (\ref{eq:mu}) is determined for a bolometric
flux, which is not directly measured. The SALT2 \citep{Guy05,Guy07} relation
(\ref{salt}) can thus be viewed as a model for a band correction,
$\DE\mu\Ns{B}$, that is linear in the variables $x\Z1$ and $c$,
\beq
\DE\mu\Ns{B}\equiv(m-M)-(m\Ns B^*-\MB)=\al x\Z1-\be c,
\eeq
where $m$ and $M$ are the bolometric apparent and absolute magnitudes
in the observer and emitter rest frames respectively.

In the SALT2 method, the light-curve parameters are simultaneously fit together
with the free cosmological parameters on the entire data set.

NGS16 assumed that all SneIa in the JLA
catalogue \citep{jla} are characterized by parameters, $\MB$, $x\Z1$ and $c$,
drawn from the same independent global Gaussian distributions,
with means $\MBn$, $x\Z{1,0}$ and $c\Z0$, and standard deviations
$\si\Z{M_{\rm B,0}}$, $\si\Z{x_{\rm 1,0}}$ and $\si\Z{c_{\rm0}}$ respectively.
These 6 free parameters were then fitted along with the light-curve
parameters $\al$, $\be$ and the cosmological parameters.

RH16 claimed that the mean light-curve stretch and colour parameters,
$x\Z{1,0}$ and $c\Z0$, of the Gaussian distributions analysed by NGS16
show some redshift dependence. This
may be partially due to astrophysical effects in the host population, or --
particularly for the colour parameter -- may arise from the
colour--luminosity relation combined with redshift--dependent detection
limits. In other words, Malmquist type biases may not be completely
corrected for in the JLA catalogue \citep{jla}. In the absence of a known
astrophysical model for such corrections, RH16
introduced 12 additional empirical parameters by replacing the global
Gaussian means according to
\beq
x\Z{1,0}\to x\Z{1,0,J} + x\Z{z,J} z,\qquad\hbox{and}\qquad c\Z0\to
c\Z{0,J} + c\Z{z,J} z,\label{RHparm}\eeq where the index $J$ runs over the four
independent subsamples in the JLA catalogue: 
(1) SNLS (SuperNova Legacy Survey); (2) SDSS (Sloan Digital Sky Survey);
(3) nearby supernovae; 
(4) HST (Hubble Space Telescope),
with $x\Z{z,4}=0$, $c\Z{z,4}=0$ on
account of limited HST data. The widths $\si\Z{x_{\rm 1,0}}$,
$\si\Z{c_{\rm0}}$ were still treated as global parameters.

\begin{figure}
\centering
\centerline{\includegraphics[scale=0.445]{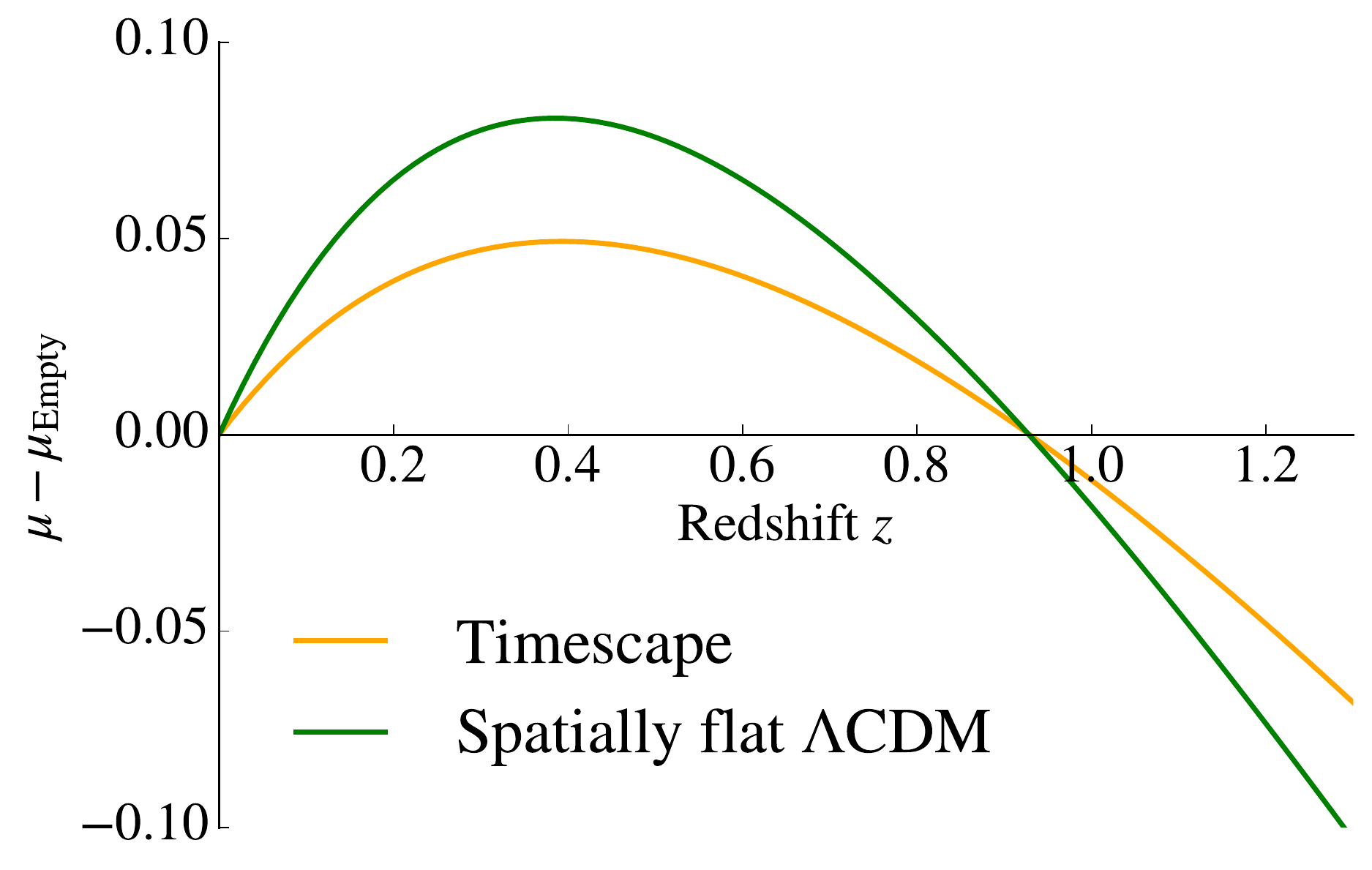}}
\caption{The residual distance moduli
$\mulcdm(z)-\mu\Ns{empty}(z)$ and $\mu\Ns{TS}(z)-\mu\Ns{empty}(z)$
with the same $\Hn$. The best-fitting parameters of Table~\ref{tab:maxL}
are assumed: $\OmMn=0.365$ for spatially flat \LCDM\ and $\fvn=0.778$ for \ts.
For redshifts $z<0.927$ over which $\mu\Ns{TS}(z)<\mulcdm(z)$, the
maximum difference between the \LCDM\ and \ts\ distance moduli is
$\mulcdm(z)-\mu\Ns{TS}(z)=0.031$ at $z=0.372$.}
\label{mures}
\end{figure}

We perform a Bayesian comparison of fits of the JLA catalogue \citep{jla} to the
luminosity distance-redshift relation for the spatially flat \LCDM\ model,
and for the \ts\ model \citep{clocks,sol,obs,dnw}. We first use the MLE
methodology of NGS16 directly,
and then investigate the effect of changes to light-curve fitting suggested by
RH16. An empty universe with constant expansion rate is
also analysed, as a convenient demarcation of accelerating from
non-accelerating expansion in the FLRW case.

Details of the theoretical luminosity distances used in (\ref{eq:mu}) are
given in Appendix \ref{lumd}. The model differences that we are testing are
best appreciated by comparing the distance moduli of the \ts\ and spatially
flat \LCDM\ models relative to an empty universe, as shown in Fig.~\ref{mures}.
The \ts\ distance modulus, $\mu\Ns{TS}(z)$, is closer to \LCDM\ than the
empty case. Nonetheless, $\mu\Ns{TS}(z)$ is always closer to $\mu\Ns{empty}(z)$
than $\mulcdm(z)$ is, a consequence of cosmic acceleration being an apparent
effect in the \ts\ model.

Further technicalities about systematic issues in implementing the
SALT2 method are discussed in Appendix \ref{salt2}.

\section{Statistical methods}
\subsection{The likelihood construction}\label{likelihoodconstruction}
We adopt the likelihood construction \citep{mtbsv} used by NGS16. The
likelihood, $\mathcal{L}$, is the probability density of the observed data --
here $(\hat{z},\hmb,\hat{x}\Z1,\hat{c})_i$, $i=1,2,\dots,N$ on $N$ supernovae
-- given a model, $\Mmod$. The likelihood can be written as \citep{mtbsv}
\begin{align} \label{expandedlikelihood} \mathcal{L}&\equiv
\mathcal{P}\left[(\hat{z}, \hmb , \hat{x}\Z1, \hat{c})_i \,|\, \Mmod\right]
\nonumber\\ &= \int\dd \MB ^{N}\,\dd x\Z1 ^{N}\,\dd c ^{N}\,\mathcal{P} \left[(\hat{z}, \hmb , \hat{x}\Z1, \hat{c})_i \,|\, (\MB , x\Z1 , c)_i,\Mmod \right]
\nonumber\\ &\hblank{30}\times \mathcal{P}\left[ (\MB , x\Z1 , c)_i \,|\, \Mmod\right] ,
\end{align}
where hatted quantities denote measured data values including all
experimental noise, and unhatted quantities are intrinsic
parameters that characterize the statistical distributions from which
the supernovae are drawn. Only the intrinsic parameters satisfy the SALT2
relation (\ref{salt}). The empirical light-curve model (\ref{salt}) and the
theoretical distance modulus (\ref{eq:mu}) together constitute the model,
$\Mmod$.

\begin{table*}
\centering
\begin{tabular}{lll}
\hline
Cosmological parameter & Prior distribution & Range \\
\hline
Timescape: $\fvn$ 			& Uniform & $[0.500, 0.799]$ \\[3pt]
Flat \LCDM: $\OmMn=1-\OmLn$ & Uniform & $[0.143, 0.487]$ \\[3pt]
\hline
Nuisance parameters & Prior distribution & Range \\
\hline
$\al$       & Uniform 						 & $[0, 1]$				\\[3pt]
$x\Z{1,0}$    & Uniform 						 & $[-20, 20]$		\\[3pt]
$\si\Z{x_{1,0}}$ & Uniform on $\log\Z{10}\si\Z{x_{1,0}}$ & $[-5, 2]$	\\[3pt]
$\be$       & Uniform 						 & $[0, 4]$				\\[3pt]
$c\Z{0}$     & Uniform 						 & $[-20, 20]$ 			\\[3pt]
$\si\Z{c_0}$   & Uniform on $\log\Z{10}\si\Z{c_0}$ 	 & $[-5, 2]$ 		\\[3pt]
$\MBn$      & Uniform 						 & $[-20.3, -18.3]$			\\[3pt]
$\siMB$      & Uniform on $\log\Z{10}\siMB$ 			 & $[-5, 2]$ 			\\
\\ \noalign{\vskip-5pt}
Additional stretch and colour	&	\multirow{2}{*}{Uniform} & \multirow{2}{*}{$[-20,20]$} \\
parameters for models II-VIII	&& 														\\
\hline
\end{tabular}
\caption{All nuisance parameters in each model have identical priors.
In the \ts\ model $\OmMn$ is defined in terms of $\fvn$ hence we
take the latter to be the more `fundamental' parameter and assign
the prior to it.}
\label{table:priors}
\end{table*}
The expansion in (\ref{expandedlikelihood}) allows one to explicitly model
the intrinsic scatter of the supernovae.
For the NGS16 model (I) we assume that the intrinsic parameters of
each supernova are drawn from identical independent Gaussian
distributions
\begin{align} &\mathcal{P}[ (\MB , x\Z1 , c)_i \,|\, \Mmod ]
=\prod_i^N \mathcal{P}[ (M\Z{{\rm B},i}, x\Z{1,i}, c\Z{i}) \,|\,\Mmod]
\nonumber\\ &\hblank{3}= \prod_i^N \mathcal{N}(M\Z{{\rm B},i}\,;\, \MBn,\sigma\Z{M_{{\rm B},0}})
\nonumber\\ &\hblank{15}\times
\mathcal{N}(x\Z{1,i}\,;\, x\Z{1,0}, \sigma_{x\X{1,0}})\;
\mathcal{N}(c\Z{i}\,;\, c\Z0, \sigma_{c\X0}),\label{ngsP}
\end{align}
where $\mathcal{N}(y\,;\,y\Z0,\si)\equiv(2\pi\si^2)^{-1/2}\exp\left[-(y-y\Z0)^2
/(2\si^2)\right]$ for each triple $\{y,y\Z0,\si\}$, with $3N\times 3N$ diagonal
covariance matrix
$\Sigma_l= \mathrm{diag}\left(\si^2\Z{M\X{{\rm B},0}},\si^2_{x_{1,0}},
\si^2\Z{c\X0},\si^2\Z{M\X{{\rm B},0}} , \dots\right)$.
The experimental part of the likelihood (\ref{expandedlikelihood}),
$\mathcal{P}\left[(\hat{z},\hmb,\hat{x}\Z1,\hat{c})_i \,|\, (\MB,x\Z1,c)_i,
\Mmod\right]$ is assumed to be a Gaussian in the intrinsic supernova
parameters, and the covariance matrix of experimental statistical and
systematic uncertainties is denoted $\Sigma_d$.
Performing the integral in (\ref{expandedlikelihood}) one obtains the final expression of the likelihood \citep{ngs}
\begin{align} \label{eq:likefin}
\mathcal{L}&=\mathcal{P}\left[(\hat{z}, \hmb , \hat{x}\Z1, \hat{c})_i \,|\, \Mmod\right]
= \mathcal{P}\left[ ( \hmb - \mu(\hat{z}) , \hat{x}\Z1 , \hat{c} )_i \,|\, \Mmod\right]
\nonumber \\
&= \left[(2\pi)^{3N}\det\left(\Sigma_d+A^\intercal\Sigma_l A\right)
\right]^{-1/2}
\nonumber\\&\hblank{5}\times\exp\left[-\frn12\left(\hat Z-Y\Z0 A\right)
\left(\Sigma_d+A^\intercal\Sigma_l A\right)^{-1}\left(\hat Z-Y\Z0 A\right)
^\intercal\,\right]
\end{align}
where $\hat Z=({\hat m}\Z{{\rm B},1}^*-\mu(\hat z\Z1),\hat{x}\Z{1,1},
\hat{c}\Z1,{\hat m}\Z{{\rm B},2}^*-\mu(\hat z\Z2),\dots)$ is a $3N$-dimensional
row vector containing the distance modulus residual and light-curve data,
$Y\Z0=(\MBn,\,x\Z{1,0},\,c\Z0,\,\MBn,\dots)$ is a $3N$-dimensional row vector
of the intrinsic
Gaussian means, and $A$ is the block diagonal matrix that propagates $Y\Z0$
to $$Y\Z0 A=(\MBn-\alpha x\Z{1,0} + \beta c\Z0,\,x\Z{1,0},\,c\Z0,\,\dots).$$
Note that the cosmological model enters only explicitly through the conversion
$\mu(\hat{z})$ of the observed redshift to a distance modulus.
There can, however, be implicit model dependence in the experimental
covariance matrix\footnote{The propagation of the error $\si_z$ to $\si_\mu$
depends on the model. However, by (\ref{muTS})--(\ref{muempty}), to leading
order for small $z$, $\si_\mu\simeq5\si_z/(z\,\ln10)$ for all cases.}
$\Sigma_d$ or in corrections made to data prior to the analysis.

To implement the RH16 parametrization (\ref{RHparm}) we replace
(\ref{ngsP}) by
\begin{align} &\mathcal{P}[ (\MB , x\Z1 , c)_i \,|\, \Mmod ]
\nonumber\\ &\hblank{3}
= \prod_{J=1}^4\;\prod_{i=1}^{N\X J} \mathcal{N}(M\Z{{\rm B},i},\,\MBn,\,
\si\Z{M_{{\rm B},0}})
\nonumber\\ &\hblank{4.5}\times
 \mathcal{N}(x\Z{1,i},\,x\Z{1,0,J} + x\Z{z,J}\hat z,\,
\si_{x\X{1,0}})\;\mathcal{N}(c\Z{i},\,c\Z{0,J}+c\Z{z,J}\hat z,\,\si_{c\X0}).
\end{align}
We recover (\ref{eq:likefin}) with the one difference: in place of
three repeated entries, the vector $Y\Z0$ is now partitioned
into different pieces for each subsample,\break $Y\Z0=(\MBn,\,x\Z{1,0,1}+
x\Z{z,1}{\hat z}\Z1,\,c\Z{0,1} + c\Z{z,1}{\hat z}\Z1,\,\dots,\,
\MBn,\,x\Z{1,0,2}+x\Z{z,2}{\hat z}\Z i,\,c\Z{0,2} + c\Z{z,2}{\hat z}\Z i,\dots,
\dots,\MBn,x\Z{1,0,4},c\Z{0,4})$.

From the likelihood (\ref{eq:likefin}) we can define frequentist confidence
regions and goodness of fit measures or alternative Bayesian versions of
these, following conventional statistical procedures summarized in Appendix
\ref{stat}.

In practice, estimating the Bayesian evidence is a computationally
intensive task, much more so than what is required to obtain parameter
estimates. We use standard Markov Chain Monte Carlo (MCMC) methods to sample
parameter space. We estimate the evidence using the publicly available
{\sc MultiNest} \citep{multinest} code,\footnote{This package is based on the
Nested sampling algorithm \citep{SkillingNS}.} with Python interface
PyMultinest \citep{pymultinest}, for the efficient evaluation of
the evidence integral \eqref{eq:evidence} with likelihood (\ref{eq:likefin}).
The accuracy of the Bayesian evidence estimate is controlled by the
number of `live' points, $n\Ns{live}$, with an error
$\sigma\sim O(n^{-1/2}\Ns{live})$.
In our analysis we choose 1000 points for the 8 or 9 parameter base model
and add 100 more points for each additional parameter.

\subsection{Choice of priors}
Given the sensitivity of the Bayes factor to priors it is important these are
chosen as objectively as possible. The choice of priors are summarized in Table
\ref{table:priors}.

\subsubsection{Nuisance parameters}
All nuisance parameters are common to both \ts\ and \LCDM\ models and
we therefore assign the same priors to both models. Where
possible,\footnote{Given the complications introduced by empirical
changes (\ref{RHparm}) to $x\Z{1,0}$, $c\Z0$, we adopt uniform priors for
for these parameters.} we adopt
priors that have been used in previous Bayesian studies of the SALT2
method \citep{mtbsv,bahamas}.
The standard deviations $\{\si\Z{x_{1,0}}, \si\Z{c_0}, \siMB\}$ are `scale'
parameters (of the residuals) and so it is more appropriate
to assign a log-uniform prior to these parameters.
The priors for the nuisance parameters are wide to ensure the most likely
regions of parameter space are supported, and provided they are wide enough,
this will have no overall effect on the Bayes factor (as the evidence of
each model will be similarly scaled).

\subsubsection{Cosmological parameters}
Only one free cosmological parameter can be constrained by supernovae:
$\OmMn$ for spatially flat \LCDM\ or $\fvn$ for the \ts\ model. Conventionally,
the combination of $\OmMn$ and $\Hn$ for the standard cosmology is strongly
constrained by the CMB acoustic peaks \citep{planck}. Measurements of the
Baryon Acoustic Oscillation (BAO) scale in galaxy clustering
statistics \citep{BOSS1,BOSS2} at low redshifts and the Lyman $\alpha$
forest \citep{LymanA,BOSS1} provide independent constraints. In the case of the
\ts\ model, however, our ability to model the CMB is still limited by
systematic uncertainties of 8--13\% \citep{nw}.

\begin{table*}
\begin{tabular}{clcccccccc}
\hline
Model &\hfil $\OmMn$\hfil
& $\al$ & $x\Z{1,0}$ & $\sigma\Z{x_{1,0}}$ & $\be$ & $c_0$ & $\sigma_{c_0}$ & $\MBn$ & $\siMB$
\\ \hline
Timescape &$0.309^{+0.070\;(1\si)\;0.127\;(2\si)}_{-0.088\;(1\si)\;0.210\;(2\si)}$
& 0.134 & 0.1050 & 0.899 & 3.13 & -0.0211 & 0.0689 & -19.1 & 0.104
\\ \noalign{\smallskip}
Spatially flat $\Lambda$CDM &
$0.365^{+0.033\;(1\si)\;0.066\;(2\si)}_{-0.031\;(1\si)\;0.060\;(2\si)}$
& 0.134 & 0.1061 & 0.899 & 3.14 & -0.0215
& 0.0688 & -19.0 & 0.104
\\
Empty universe & $-$
& 0.133 & 0.1013 & 0.900 & 3.13
& -0.0204 & 0.0690 & -19.0 & 0.106
\\
\hline
\end{tabular}
\caption{Best-fitting MLE parameters corresponding to the likelihood $\mathcal{L}
({\rm Data} | \Mmod) $ with the model $\Mmod$ representing the cosmological
model, the SALT2 procedure and the intrinsic distributions of SneIa parameters.
SneIa at redshifts $z<0.033$ (statistical homogeneity scale) are excluded.
Confidence limits are given for the one free cosmological parameter. In the
\ts\ case this corresponds to
$\fvn=0.778^{+0.063\;(1\si)\;0.155\;(2\si)}_{-0.056\;(1\si)\;0.104\;(2\si)}$.
The value of $\MBn$ is obtained for $h=0.668$ for the \ts, and $h=0.7$ for the
two FLRW models.
The difference of parameters from NGS16 is principally due to the SHS
cut at $\zmin=0.033$, the effect of which is seen in Fig.~\ref{lcurve}.}
\label{tab:maxL}
\end{table*}
We therefore determine priors for $\fvn$ in the \ts\ model using best present
knowledge. For the CMB we use results of a model-independent analysis of
the acoustic peaks \citep{asas} with Planck satellite
data, and choose a prior from a 95\% confidence fit of the angular scale of
the sound horizon. To date BAO studies all implicitly assume the FLRW model,
and do not yet provide an equivalent model independent constraint. We
therefore adopt a prior using FLRW-model estimates of the angular diameter of
the BAO scale, including the full range of values which are currently in
tension \citep{LymanA,BOSS1,BOSS2}. We take generous 95\% confidence
limits determined by assuming that {\em either} the low redshift galaxy
clustering results \citep{BOSS1,BOSS2} {\em or} the $z=2.34$ Lyman-$\alpha$
results \citep{LymanA} are correct. Priors for the spatially flat \LCDM\
model are determined by an identical methodology. Further details are given
in Appendix \ref{prior}.

\section{Results}

\subsection{Analysis with supernova parameters drawn from global Gaussian
distributions}

Since there is a degeneracy
between the Hubble constant, $\Hn$, and the magnitude, $\MBn$, we fix
$\Hn$ for each model. The value of $\MBn$ then depends on this choice.
We are then left with one free cosmological parameter, the matter density
parameter $\OmMn$ in the spatially flat \LCDM\ model, and the present epoch
void fraction $\fvn$ in the \ts\ model. We can alternatively define an
effective ``dressed matter density parameter'' $\OmMn=\frn12(1-\fvn)(2+\fvn)$
\citep{clocks,obs}, which takes similar numerical values to the concordance
\LCDM\ model, allowing likelihood functions to be plotted on the same scale.
(This parameter does not obey the Friedmann equation sum rule, however.) The
9 parameters $\{\OmMn,\al,x\Z{1,0},\si\Z{x_{\rm 1,0}},\be,c\Z0,\si\Z{c_{\rm0}}
,\MBn,\si\Z{M_{\rm B,0}}\}$ are then fit for each model by determining the
likelihood function with all parameters other than $\OmMn$ (or $\fvn$)
treated as nuisance parameters. The empty universe has 8
parameters since $\OmMn=0$.

\subsubsection{Statistical homogeneity scale cuts}
An important systematic issue in the \ts\ cosmology is the fact that
an average expansion law only holds on scales greater than the statistical
homogeneity scale (SHS) $\gsim70$--$120\hm$ \citep{h05,sdb12}. This corresponds
to a CMB rest frame redshift of order $z\goesas0.023$--$0.04$. In fact, SneIa
analyses using the MLCS method have typically excluded SneIa below
a cutoff at $z=0.024$ \citep{R07}. However,
the JLA catalogue \citep{jla} includes $53$ SneIa, with $z<0.024$.

Following \citet{sw} we determine cosmological model
distances in the CMB frame, but make a redshift cut at the SHS, taken at
$\goesas100\hm$. Furthermore, to examine the effect of the
SHS cut on the fit of light-curve parameters, we perform the entire analysis
while progressively varying the minimum redshift in the range $0.01\le\zmin
< 0.1$; i.e., up to a redshift 3 times larger than the SHS. Systematic
effects associated with the SHS can then be revealed. Our key results
will be quoted for a cut at $z\Ns{SHS}=0.033$ in the CMB rest frame.
The best-fitting MLE parameters
with $\zmin=0.033$ are presented in Table~\ref{tab:maxL}.

\begin{figure*}
\centering
\centerline{\includegraphics[scale=0.43]{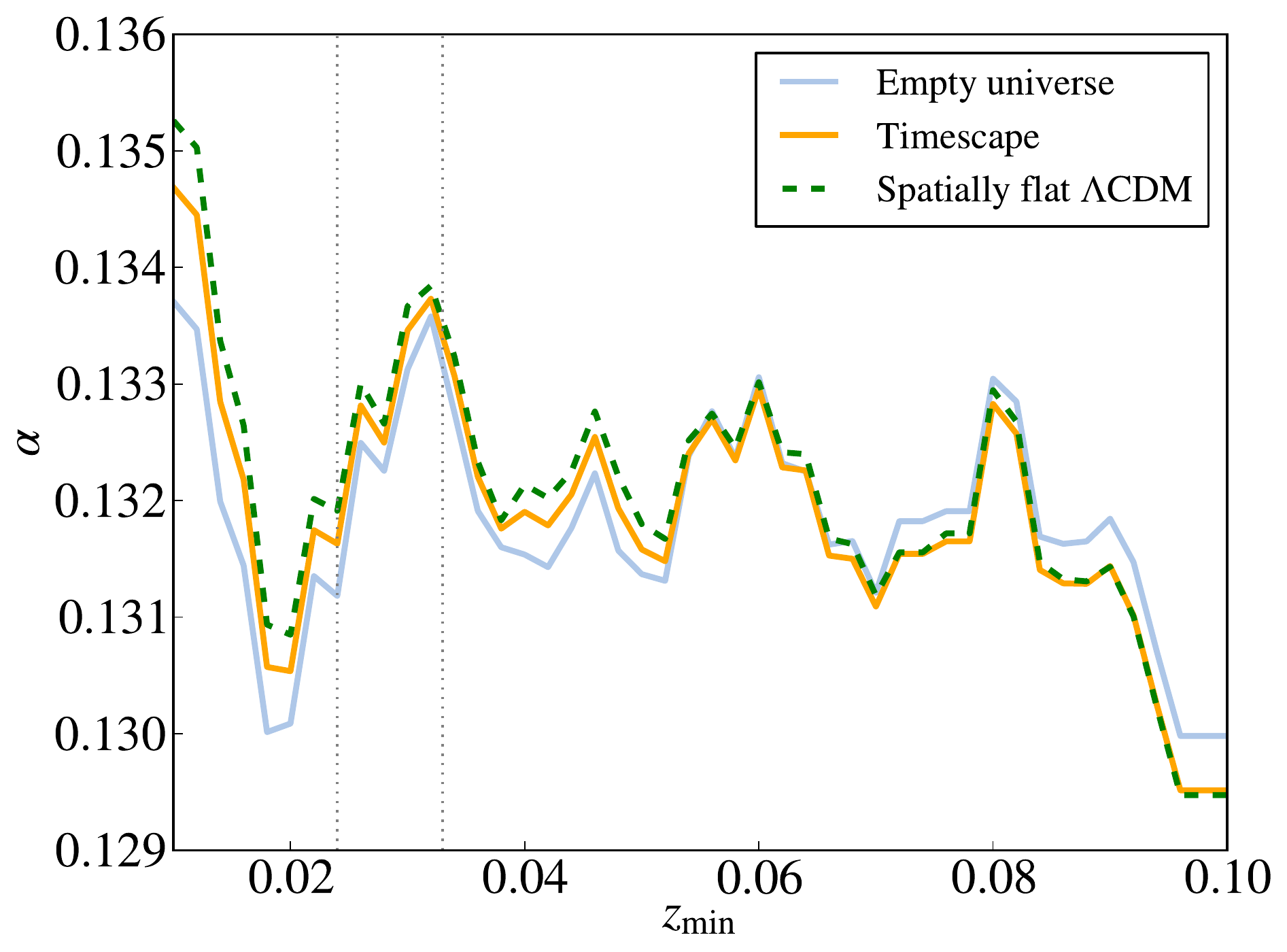}
\includegraphics[scale=0.43]{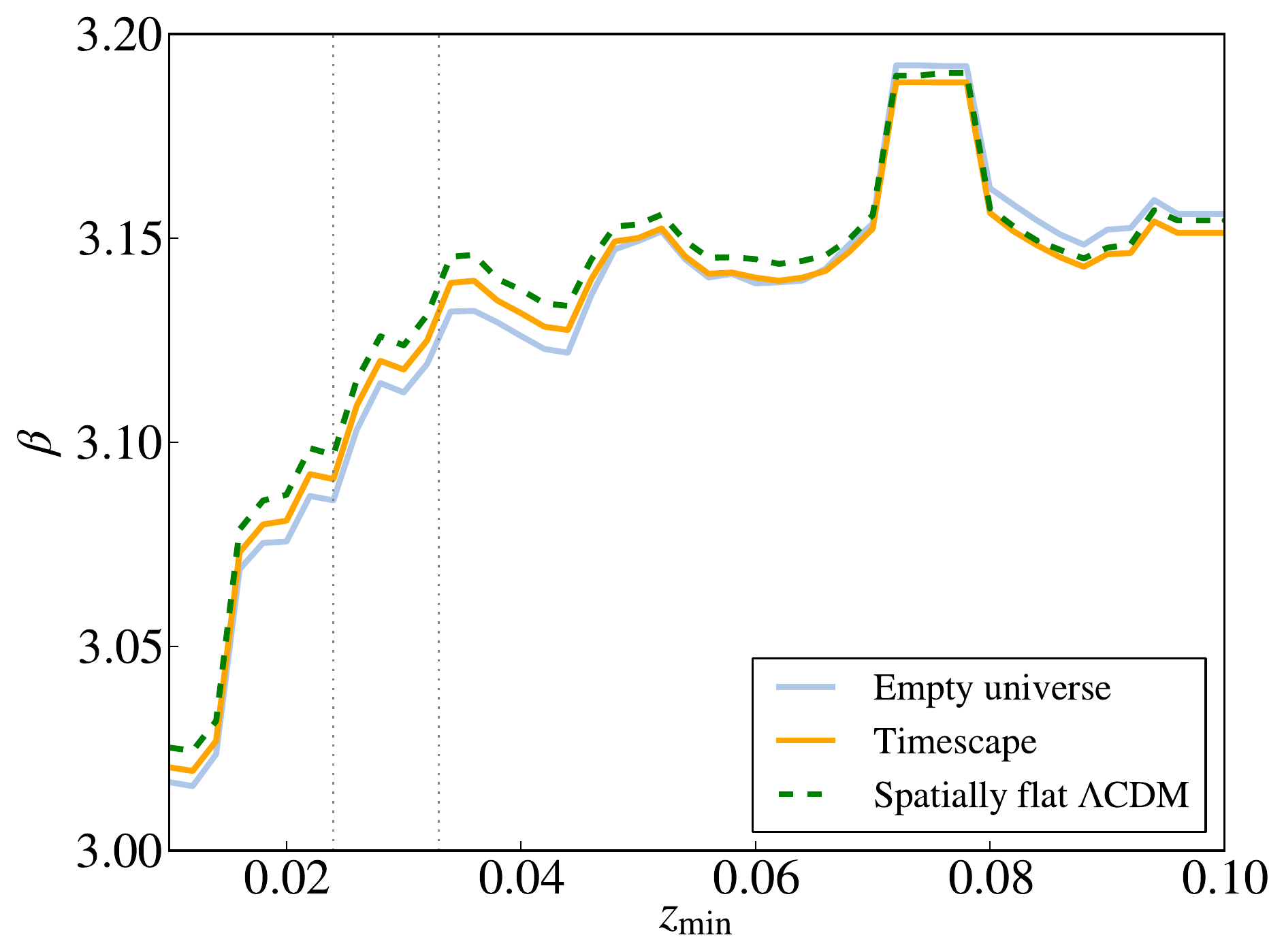}}
\vskip-10pt\leftline{\hblank{10}{\bf(a)}\hblank{80}{\bf(b)}}\bigskip
\centerline{\includegraphics[scale=0.43]{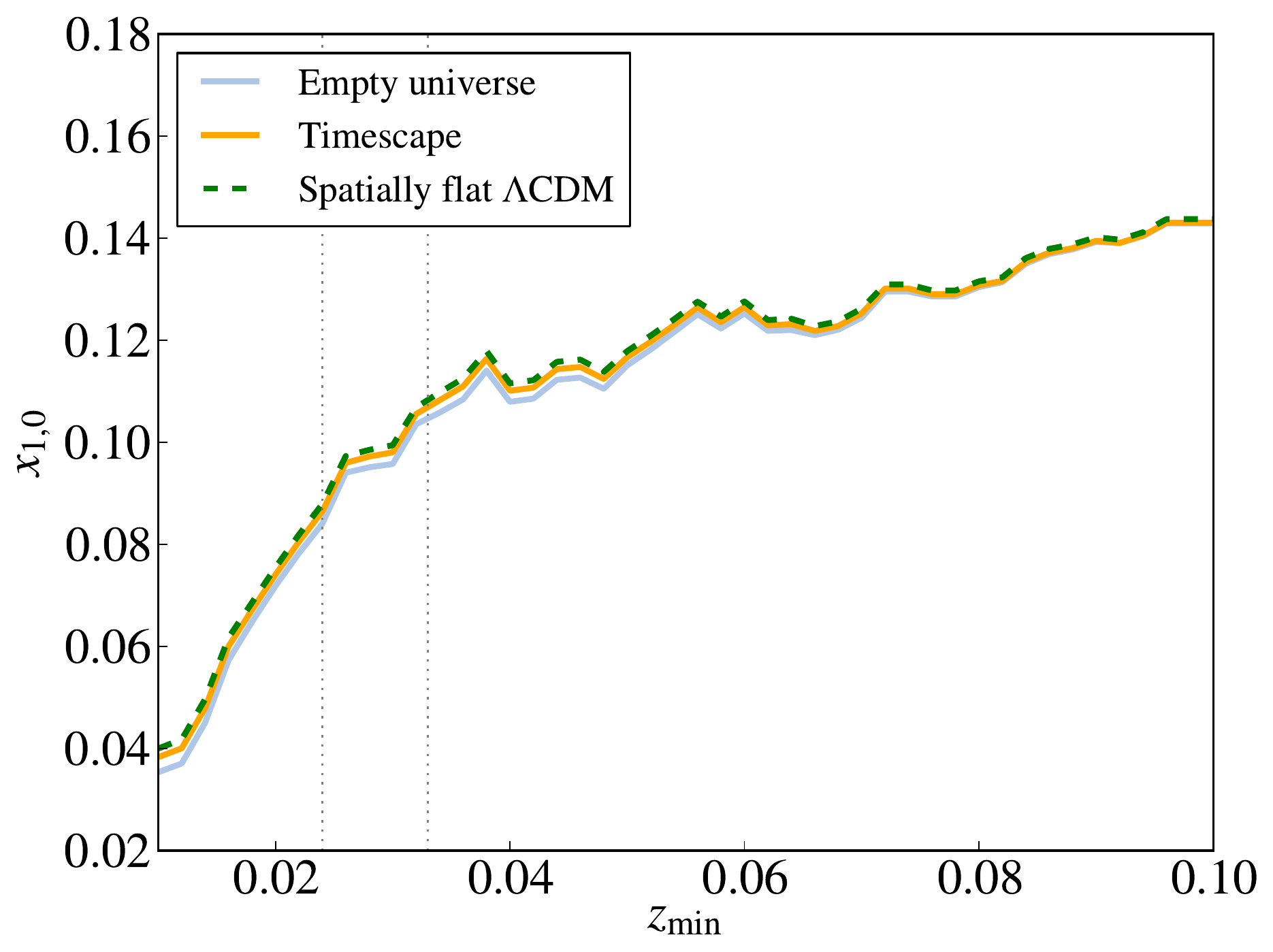}
\includegraphics[scale=0.43]{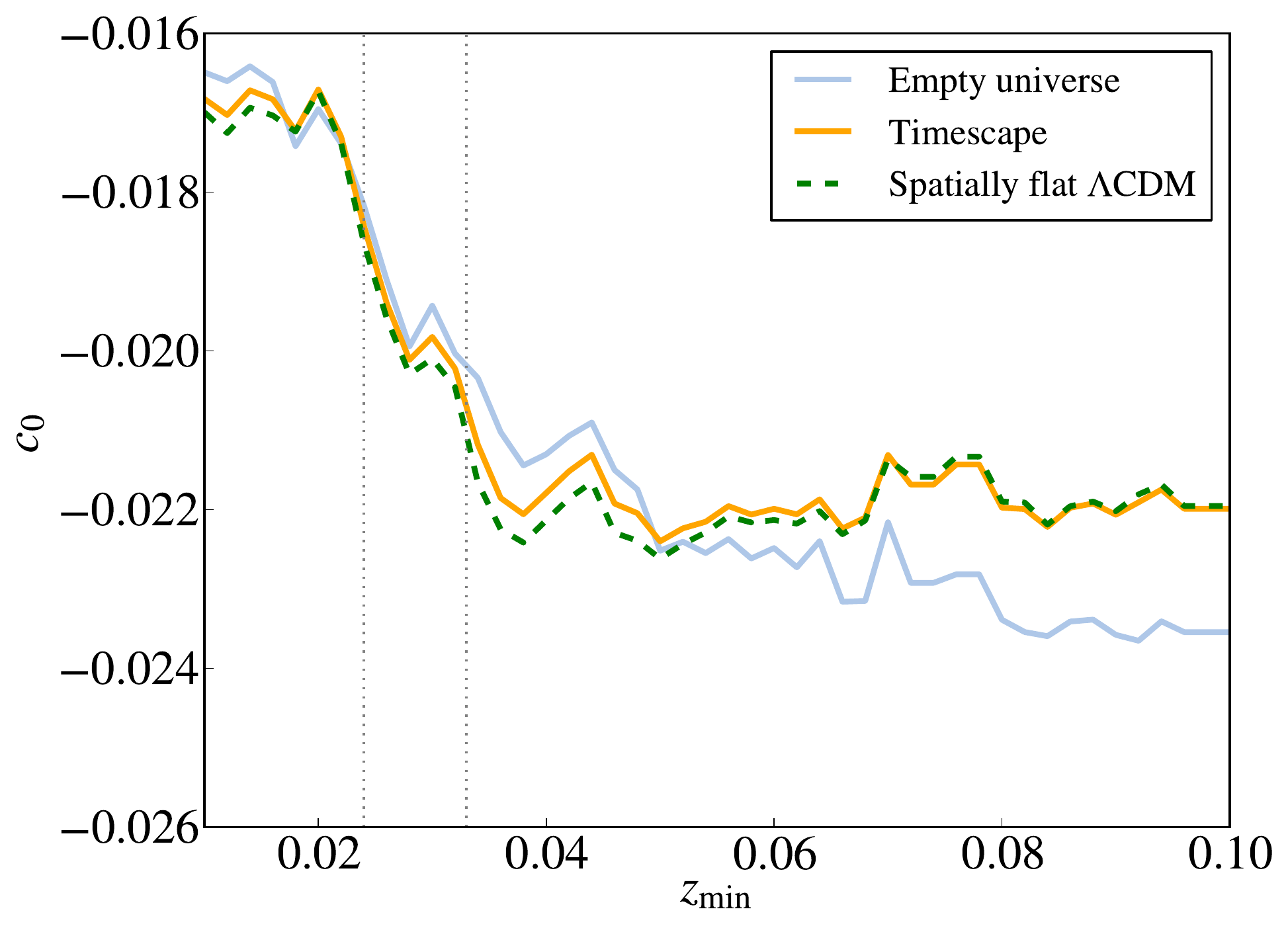}}
\vskip-10pt\leftline{\hblank{9}{\bf(c)}\hblank{82}{\bf(d)}}\bigskip
\centerline{\includegraphics[scale=0.555]{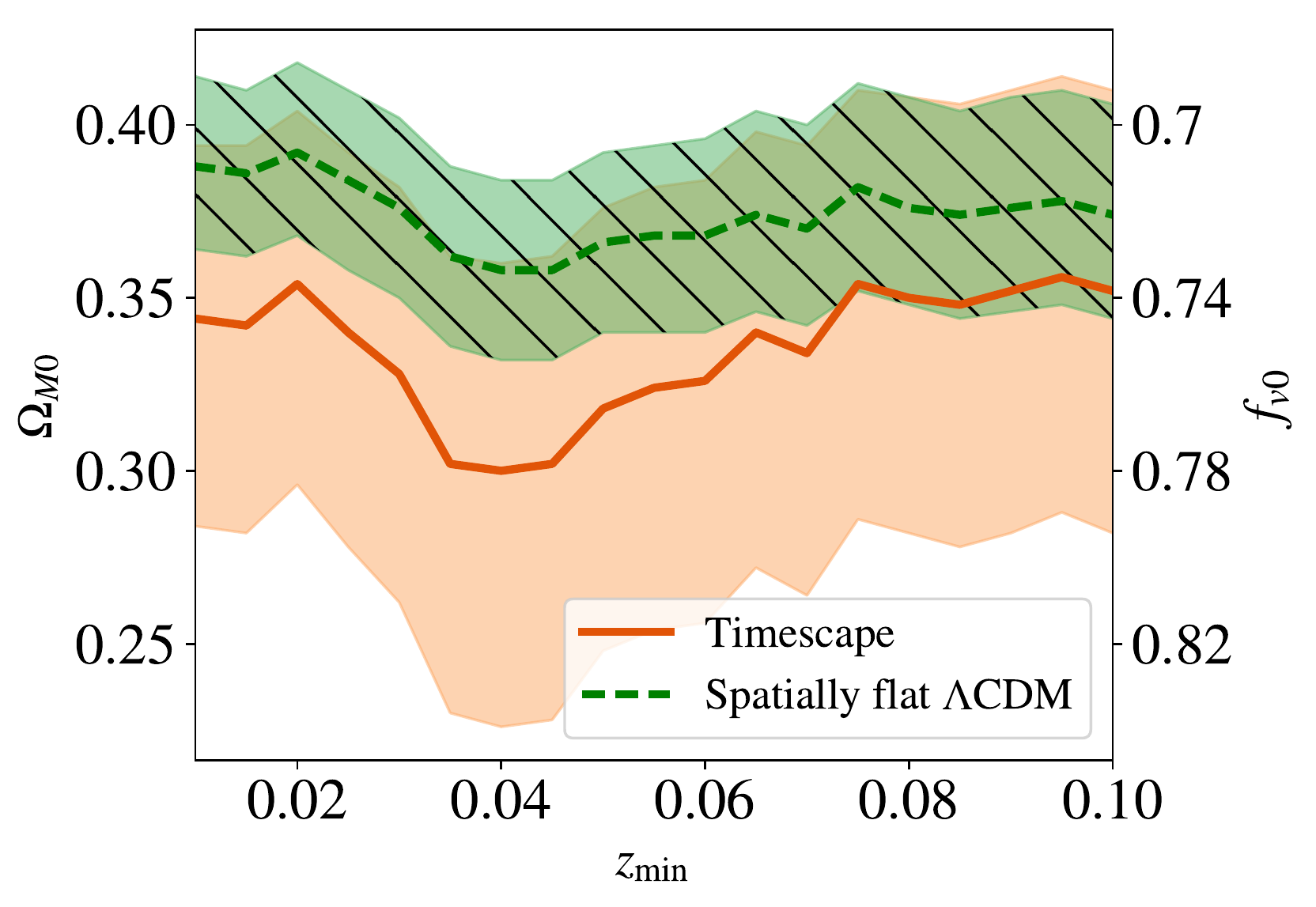}
\includegraphics[scale=0.44]{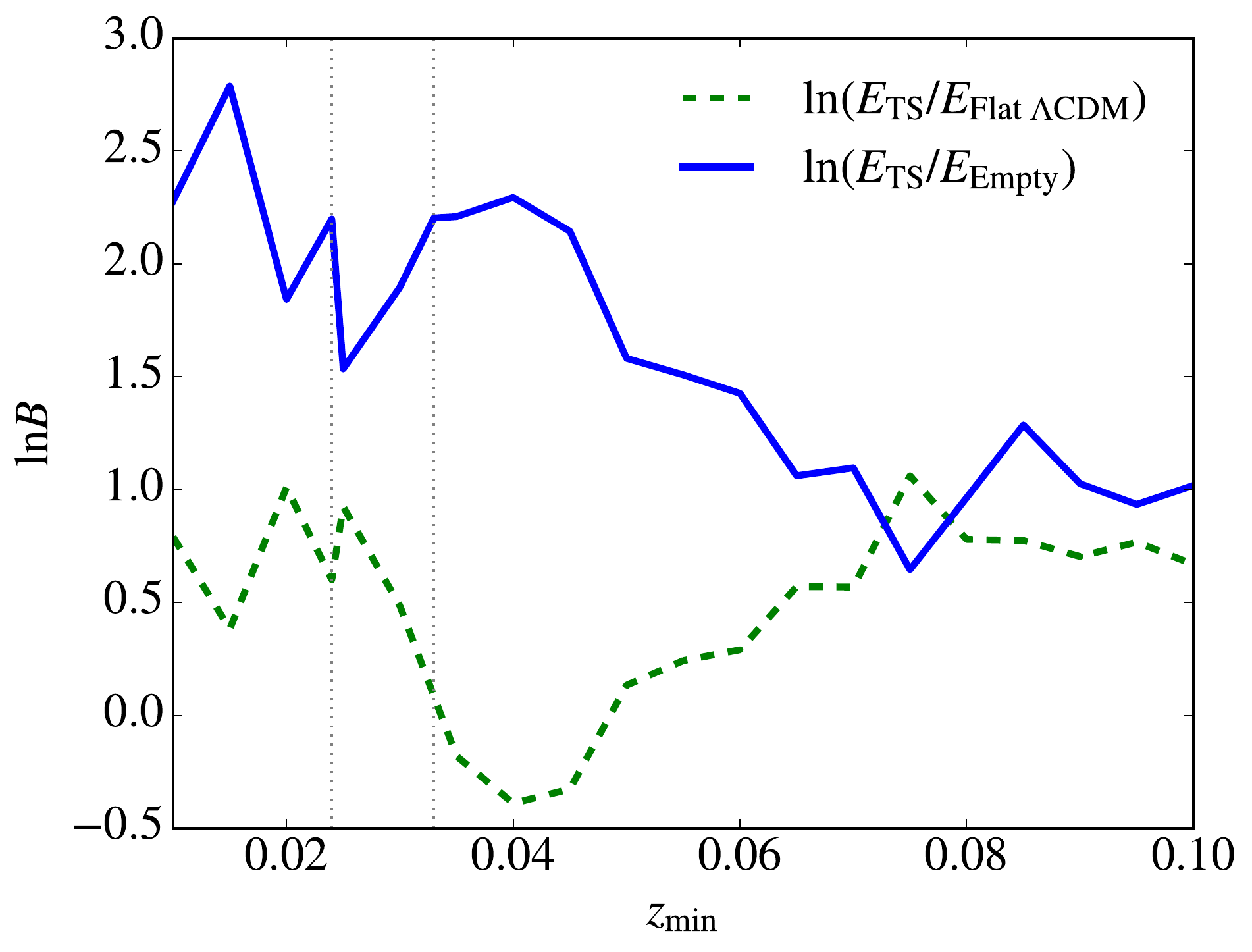}}
\vskip-10pt\leftline{\hblank{9}{\bf(e)}\hblank{82}{\bf(f)}}
\caption{MLE best-fitting parameters, and Bayes factor, for the NGS16 model
as $\zmin$
is varied: {\bf(a)} $\al$; {\bf(b)} $\be$; {\bf(c)} $x\Z{1,0}$; {\bf(d)} $c\Z0$;
{\bf(e)} $\OmMn$ (or $\fvn$) with $1\si$ bounds;
{\bf(f)} $\ln B=\ln(E\Ns{TS}/E\ns{model})$.
Vertical dotted lines at $\zmin=0.024$ and $\zmin=0.033$ indicate the expected
rough redshift range of an emerging statistical homogeneity scale.}
\label{lcurve}
\end{figure*}

For the priors given in Table \ref{table:priors} the Bayesian evidence in
favour of the \ts\ model relative to the spatially flat \LCDM\ model
is $\ln B=0.085\pm0.01$ with $\zmin=0.033$, or
$\ln B=0.600\pm0.007$ with
$\zmin=0.024$. Since $|\ln B|<1$ the two models\footnote{Both models have
positive relative Bayesian evidence compared to the empty model. Although the
evidence is not particularly strong, $|\ln B\Z2|\goesas2.2$ incorporates
priors which demand standard recombination for both \LCDM\ and timescape. By
that criterion the empty model is simply ruled out.}
are statistically
indistinguishable. This conclusion is insensitive to O(1$\si$) changes to the
width of the uniform priors on $\fvn$ and $\OmMn$, or to variations of the
minimum redshift as shown in Fig.~\ref{lcurve}(e).

While the Bayes factors do not show significant variation with $\zmin$,
the values of particular best-fitting light-curve parameters show a marked
change at the SHS. As shown in
Fig.~\ref{lcurve}, there is a marked 30\% jump in $c\Z0$
as $\zmin$ is varied from $0.01$ up to $z\simeq0.033$, when compared
to the subsequent fluctuations if $\zmin$ is increased up to $0.1$. For
$x\Z{1,0}$ there is a similar jump, although a linear trend remains in the
range $0.033<\zmin<0.1$. The parameter $\be$ parameter shows a small (3\%)
jump up to the SHS followed by $\pm1$\% fluctuations, while $\al$ remains
relatively constant, fluctuating by $\pm2$\% over the whole range.

Since the light-curve parameters are remarkably close for all three cosmologies
while showing a jump as the SHS emerges (distinct from the residual $c\Z0$
trend for the empty model with $\zmin\gsim0.05$) there is clear evidence for
some systematic effect at precisely the scale where we expect it.

\begin{table*}
\begin{tabular}{lccrrrrrrrrrr}
\hline\noalign{\smallskip}
Timescape&$k$&$\OmMn$&$\ave{x\Z{1,0}}$&$\ave{c\Z{0}}$&$\ave{x\Z{z}}$&$\ave{c\Z{z}}$&BIC&$\DE\Ns{BIC}$&$\ln E$&$C_b$&$\ln B\Z1$&$\fvn$\\
\noalign{\smallskip}\hline\noalign{\smallskip}
I (NGS16)&9&0.309&0.105&-0.021&&&-185.0&&80.38&8.53&&0.778\\
II&10&0.278&-0.073&-0.021&0.511&&-199.6&-14.6&81.67&9.28&-1.28&0.802\\
III&15&0.278&-0.183&-0.021&0.806&&-169.9&15.1&66.46&14.11&13.93&0.801\\
IV&10&0.000&0.104&0.002&&-0.065&-249.9&-64.9&78.52&9.19&1.86&1.000\\
V&15&0.010&0.092&0.054&&-0.351&-157.7&27.3&81.29&14.15&-0.91&0.993\\
VI&11&0.000&-0.071&0.001&0.499&-0.062&-189.7&-4.7&79.35&10.21&1.94&1.000\\
VII (RH16)&21&0.000&-0.123&0.054&0.490&-0.348&-200.0&-15.0&65.85&19.47&14.53&1.000\\
VIII&16&0.000&-0.085&0.061&0.501&-0.348&-229.3&-44.3&82.03&15.30&-1.65&1.000\\
\hline\noalign{\smallskip}
Flat \LCDM&$k$&$\OmMn$&$\ave{x\Z{1,0}}$&$\ave{c\Z{0}}$&$\ave{x\Z{z}}$&$\ave{c\Z{z}}$&BIC&$\DE\Ns{BIC}$&$\ln E$&$C_b$&$\ln B\Z1$&$\ln B\Z2$\\
\noalign{\smallskip}\hline\noalign{\smallskip}
I (NGS16)&9&0.365&0.106&-0.021&&&-192.5&&80.30&8.93&&0.08\\
II&10&0.353&-0.069&-0.021&0.503&&-241.2&-48.7&81.64&10.01&-1.34&0.03\\
III&15&0.353&-0.186&-0.021&0.847&&-159.8&32.7&66.62&14.60&13.68&-0.16\\
IV&10&0.303&0.106&-0.002&&-0.057&-192.9&-0.4&79.60&9.98&0.70&-1.08\\
V&15&0.296&0.093&0.052&&-0.354&-228.5&-36.0&83.77&14.87&-3.47&-2.47\\
VI&11&0.292&-0.069&-0.002&0.501&-0.057&-179.1&13.4&80.87&10.97&2.89&-1.52\\
VII (RH16)&21&0.286&-0.127&0.051&0.534&-0.352&-155.3&37.2&68.97&20.58&11.33&-3.12\\
VIII&16&0.286&-0.080&0.059&0.499&-0.354&-232.8&-40.3&84.95&15.89&-4.65&-2.92\\
\noalign{\smallskip}\hline\noalign{\smallskip}
Empty&$k$&$\OmMn$&$\ave{x\Z{1,0}}$&$\ave{c\Z{0}}$&$\ave{x\Z{z}}$&$\ave{c\Z{z}}$&BIC&$\DE\Ns{BIC}$&$\ln E$&$C_b$&$\ln B\Z1$&$\ln B\Z2$\\
\noalign{\smallskip}\hline\noalign{\smallskip}
I (NGS16)&8&-&0.101&-0.020&&&-181.5&&78.18&8.11&&2.20\\
II&9&-&-0.078&-0.019&0.517&&-190.1&-8.6&79.92&9.02&-1.74&1.75\\
III&14&-&-0.095&-0.020&0.749&&-218.9&-37.4&64.43&13.75&13.76&2.03\\
IV&9&-&0.098&0.002&&-0.054&-185.7&-4.2&78.56&9.05&-0.37&-0.03\\
V&14&-&0.087&0.054&&-0.336&-180.4&1.1&79.85&14.19&-1.66&1.45\\
VI&10&-&-0.072&0.002&0.489&-0.051&-186.3&-4.8&79.62&10.17&0.23&-0.27\\
VII (RH16)&20&-&-0.122&0.054&0.460&-0.332&-198.7&-17.2&64.31&18.68&13.88&1.55\\
VIII&15&-&-0.081&0.061&0.482&-0.334&-221.4&-39.9&80.74&14.89&-2.55&1.30\\
\noalign{\smallskip}
\hline
\end{tabular}
\caption{Selected
parameters fit for $\zmin=0.033$, with the following empirical
models for light-curve parameters:
(I) constant $x\Z{1,0}$, constant $c\Z0$; (II) global linear $x\Z{1,0}$,
constant $c\Z0$; (III) split linear $x\Z{I,1,0}$, constant $c\Z0$;
(IV) constant $x\Z{1,0}$, global linear $c\Z0$;
(V) constant $x\Z{1,0}$, split linear $c\Z{0,I}$;
(VI) global linear $x\Z{1,0}$, global linear $c\Z0$;
(VII) split linear $x\Z{1,0}$, split linear $c\Z0$;
(VIII) global linear $x\Z{1,0}$, split linear $c\Z0$.\quad
Notes: $k\equiv$ number of free parameters; quantities
$\ave{\Phi}\equiv (\sum N\Z I\,\Phi\Z{I})/(\sum_{I=1}^4 N\Z I)$ denote
an average over subsamples with $I=1\dots4$ for $x\Z{1,0,I}$, $c\Z{0,I}$
and $I=1\dots3$ for $x\Z{z,I}$ , $c\Z{z,I}$ for split models or
$\ave{\Phi}\equiv \Phi$ otherwise;
BIC $=$ Bayesian Information Criterion; $E=$ Bayesian evidence;
$C_b=$ Bayesian complexity; $\DE\Ns{BIC}=\hbox{BIC}\ns{model}-
\hbox{BIC}\Ns{I}$ and $\ln B\Z1=\ln(E\Ns{I}/E\ns{model})$ are evaluated with
cosmological model fixed; $\ln B\Z2=\ln(E\Ns{TS}/E\ns{model})$ is evaluated
with light-curve model fixed.
}
\label{tab:var}
\end{table*}
\begin{figure*}
\centering
\centerline{{\bf(a)}\includegraphics[scale=0.43]{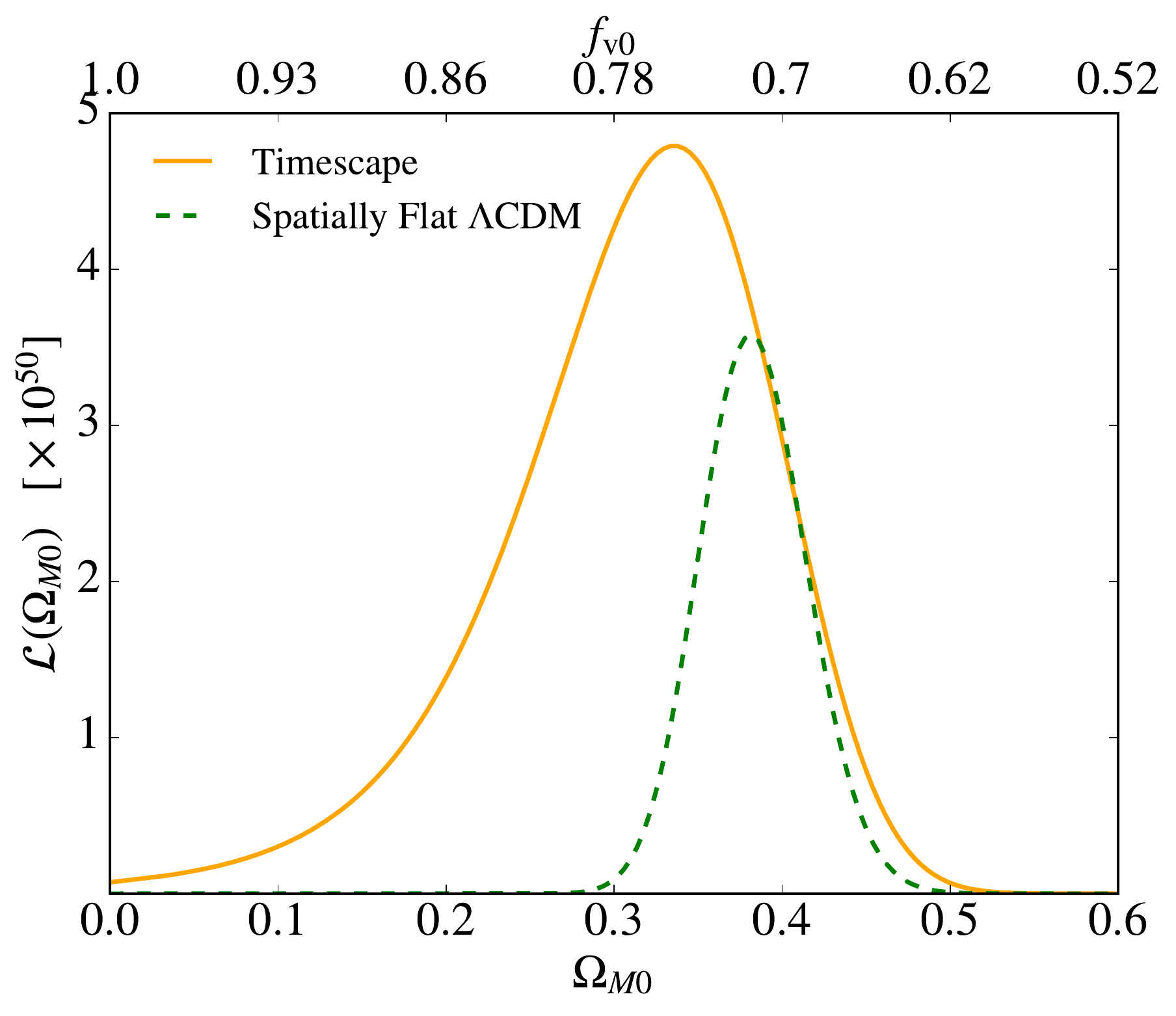}
{\bf(b)}\includegraphics[scale=0.43]{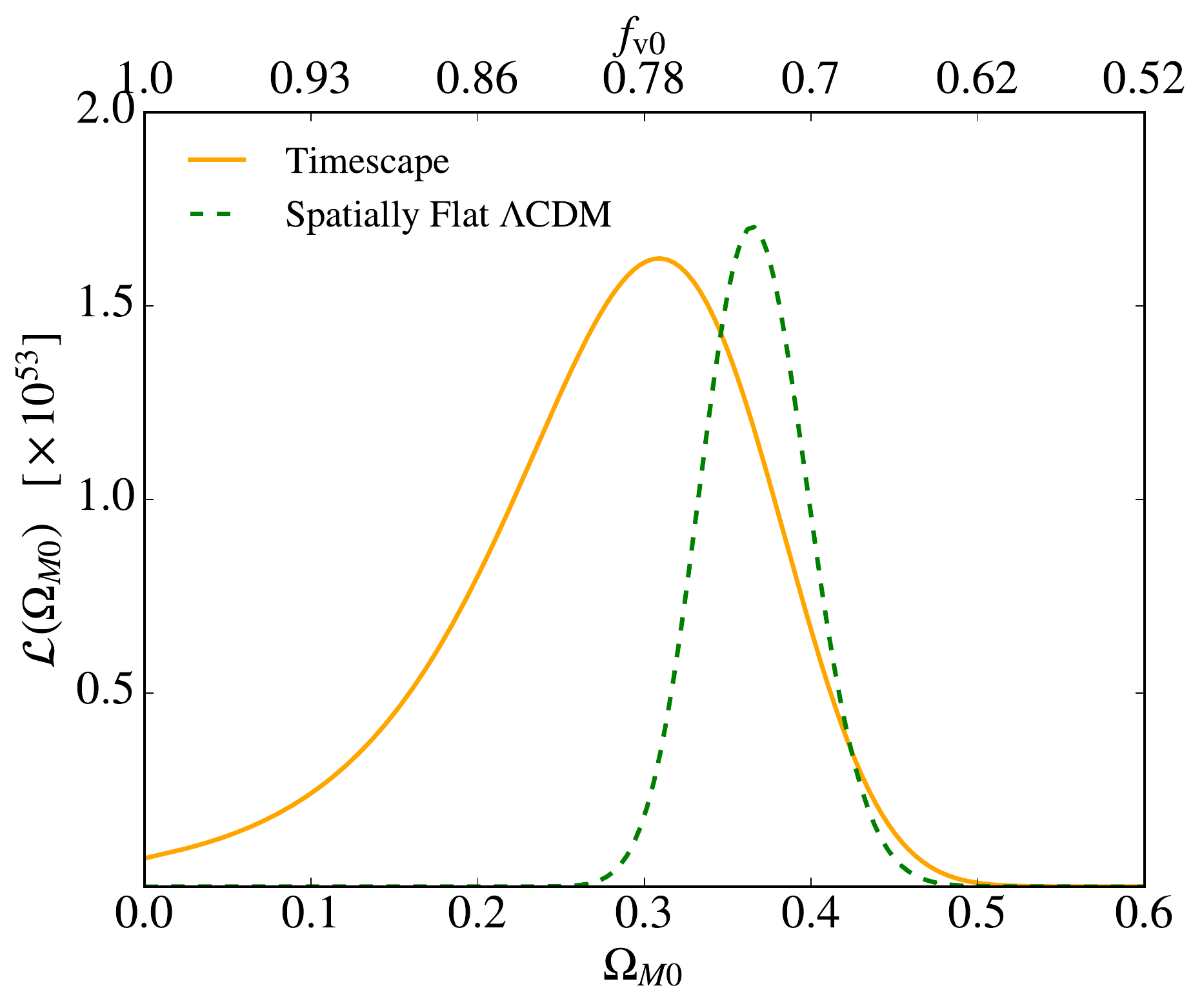}}
\centerline{{\bf(c)}\includegraphics[scale=0.43]{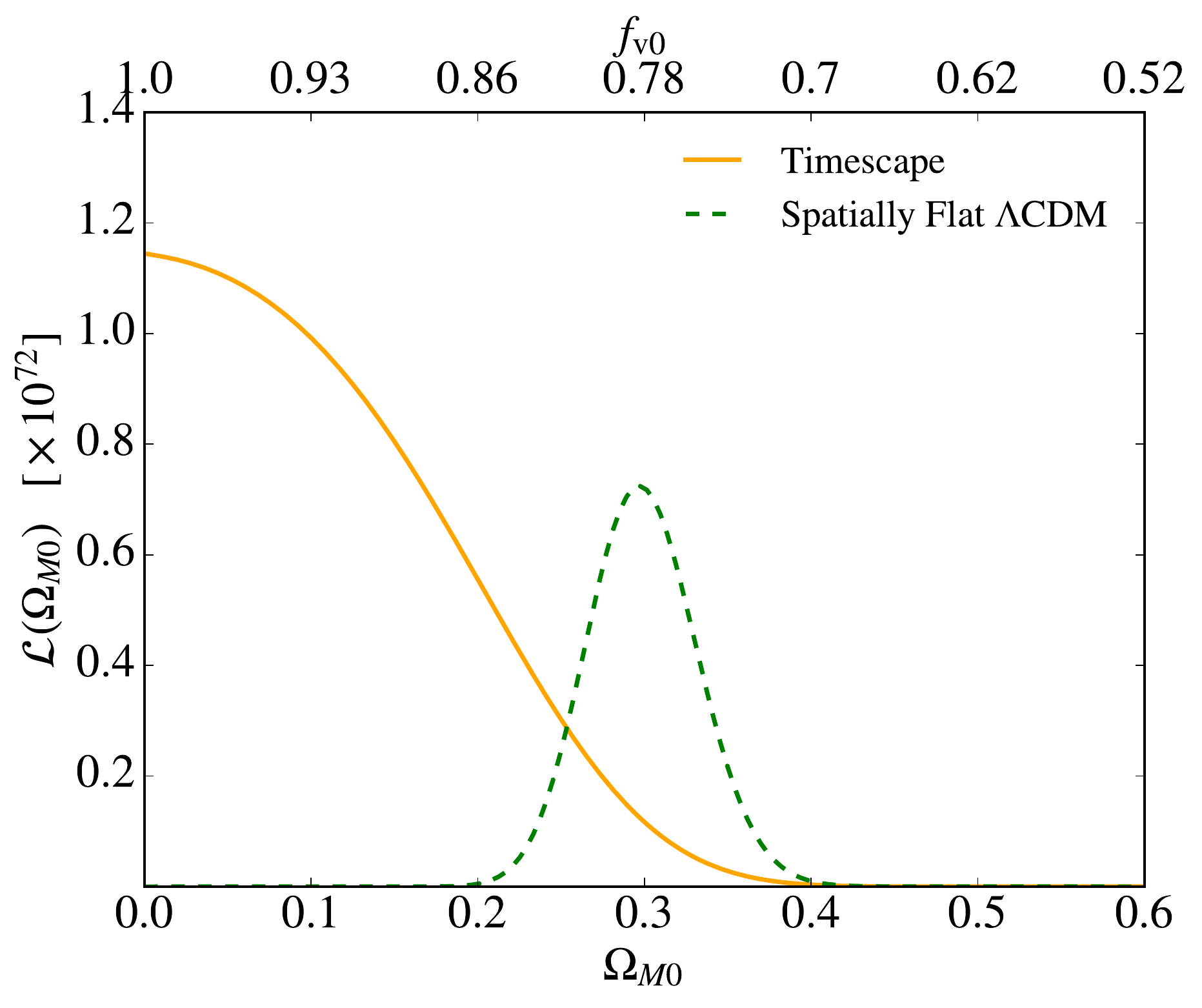}
{\bf(d)}\includegraphics[scale=0.43]{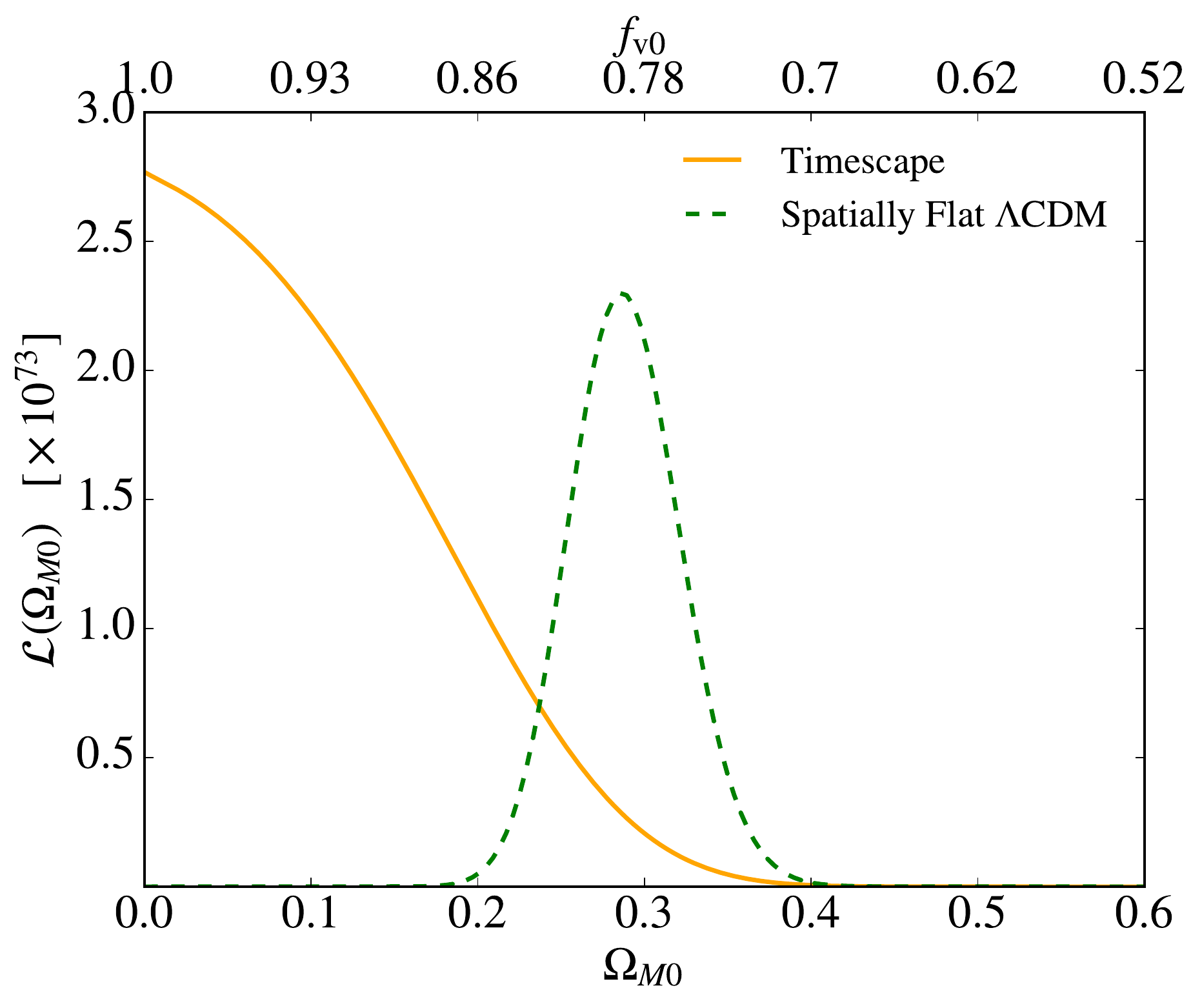}}
\caption{Profile likelihoods in $\OmMn$ and $\fvn$ for model I (NGS16) and
model VII (RH16): {\bf(a)} NGS16, $\zmin=0.024$; {\bf(b)} NGS16, $\zmin=0.033$;
{\bf(c)} RH16, $\zmin=0.024$; {\bf(d)} RH16, $\zmin=0.033$.
Model IV, V, VI and VIII results are very similar to model VII.}
\label{like4}
\end{figure*}
\subsection{Analysis with linear redshift variation for mean stretch and
colour parameters}

Although RH16 considered four distinct subsamples, the mean stretch parameter
actually shows a global increasing trend in the \LCDM\ case evident in
\citep[Fig.~1, left panels]{rh}. Our procedure of varying the minimum
redshift cut on the whole
sample also isolates any global trend. Such a trend is indeed
evident in Fig.~\ref{lcurve}(c) beyond the SHS, with $x_{1,0}$ increasing
40\% as the minimum sample redshift increases from $\zmin=0.033$ to
$\zmin=0.1$. Beyond $\zmin=0.034$ an equivalent global trend in the mean
colour parameter, $c\Z0$, is not evident in Fig.~\ref{lcurve}(d), however,
except in the case of the empty universe, which shows a 13\% decrease in
$c_0$ between $\zmin=0.034$ and $\zmin=0.1$. A global shift in $x_{1,0}$
would seem more consistent with an astrophysical systematic in the source
population, rather than sampling biases with different
thresholds for different samples.

To fully understand the differences that arise on making the RH16 changes
(\ref{RHparm}), we have also
investigated the effect of adding a smaller number of free parameters,
by considering linear $z$ relations in just one of the parameters $x\Z{1,0}$
or $c\Z0$, and the difference between global linear relations
and a split by subsamples.
The advantage of our fully Bayesian approach is that
not only can we compare the relative Bayesian
evidence for different cosmological models with the same
light-curve parameters, but we can also compare the merits of different
empirical light-curve models. The values of the Bayesian evidence are
shown in Table~\ref{tab:var}, along with a selection of parameters.
The changes to the parameters $\al$ and $\be$ are negligible
between models, and are not tabulated. We do not tabulate all additional
(up to 12) parameters for the case of the split subsamples, but an average.

\subsubsection{Stretch parameter $x\Z{1,0}$}
Consistent with remarks above, relative to the baseline NGS16 model I, light
curve model II provides positive (but not strong) Bayesian evidence for a
global linear trend in $x\Z{1,0}$ independent of cosmological model, with
$\ln B\Z1=1.28$, $1.34$, $1.74$ for the timescape, \LCDM\ and empty models
respectively. The BIC
evidence for the same conclusion is very strong (timescape, \LCDM\ models)
or strong (empty model). By contrast model III gives strong evidence
$|\ln B\Z1|>13$ against a split linear law in $x\Z{1,0}$
independent of cosmological model.
The Bayesian penalty for introducing new empirical parameters depends on the
choice of the priors, but our conclusion is robust to reasonable changes.
Furthermore, the frequentist BIC statistic
$\DE\Ns{BIC}$ also strongly disfavours model III relative to models I, II
in the \LCDM\ and \ts\ cases.

\subsubsection{Colour parameter $c\Z0$}

In contrast to the stretch parameter, results involving the colour parameter
vary greatly with cosmological model. Despite model IV having the
global minimum BIC statistic for timescape, $\ln B\Z1$ shows no
significant evidence\footnote{The empty universe
has marginal evidence, consistent with Fig.~\ref{lcurve}(d) for
$\zmin>0.05$.} for any global linear redshift law. Relative Bayesian evidence
for a split linear law in $c\Z0$ is marginal for timescape, positive for the
empty universe, and strong for \LCDM, with $\ln B\Z1=$ -0.91, -1.66 and -3.47
respectively.

The original RH16 model VII suffers similar problems to model III in terms
of Bayesian evidence, evidently on account of the split linear law in
$x\Z{1,0}$. However, model VIII has the strongest Bayesian evidence of all
models. It adds a
global linear redshift law in $x\Z{1,0}$ to model V. The improvement in
Bayesian evidence for model VIII relative to model V is marginal for \ts\ and
the empty universe, and positive for \LCDM, with $\DE\ln B\Z1=-0.74$, $-0.89$
and $-1.18$ respectively.

Although $\ln B\Z2$ for model VIII gives positive (but not strong) relative
evidence for \LCDM\ over timescape, any conclusion drawn from this depends
on additional empirical light-curve parameters which now depend on the
cosmological model\footnote{Some of the largest differences
occur in the SNLS subsample: $c\Ns{0,\lcdm}=0.0483$ and $c\Ns{0,TS}=0.0565$,
a 17\% difference. For the NGS16 model, by contrast, differences are 2\%.}.
Furthermore, since the maximum likelihoods are comparable, the difference
in Bayesian evidence is primarily due to the \ts\ maximum likelihood being
driven to the unphysical limit $\fvn\to1$ for any light-curve model with
linear variations in $c\Z0$, as is shown in Fig.~\ref{like4}, which
compares likelihoods in $\OmMn$ (or $\fvn$) for the NGS16 and RH16 models for
two choices of $\zmin$.

Very similar results were found by \citet{sw} in applying
SALT2 without the methodology of NGS16, leading to a large discrepancy
in the predictions of the SALT2 and MLCS2k2 fitters for \ts.
Since direct application of the NGS16 methodology to the JLA catalogue
agrees with some previous MLCS2k2 fits to smaller data sets \citep{lnw,sw},
we conclude that systematics similar to linear redshift variations in $c\Z0$
may be the key to earlier discrepancies.

\subsubsection{Cosmological model dependency of linear redshift changes to
SALT2 methodology}
\begin{figure*}
\centering
\centerline{\includegraphics[scale=0.35]{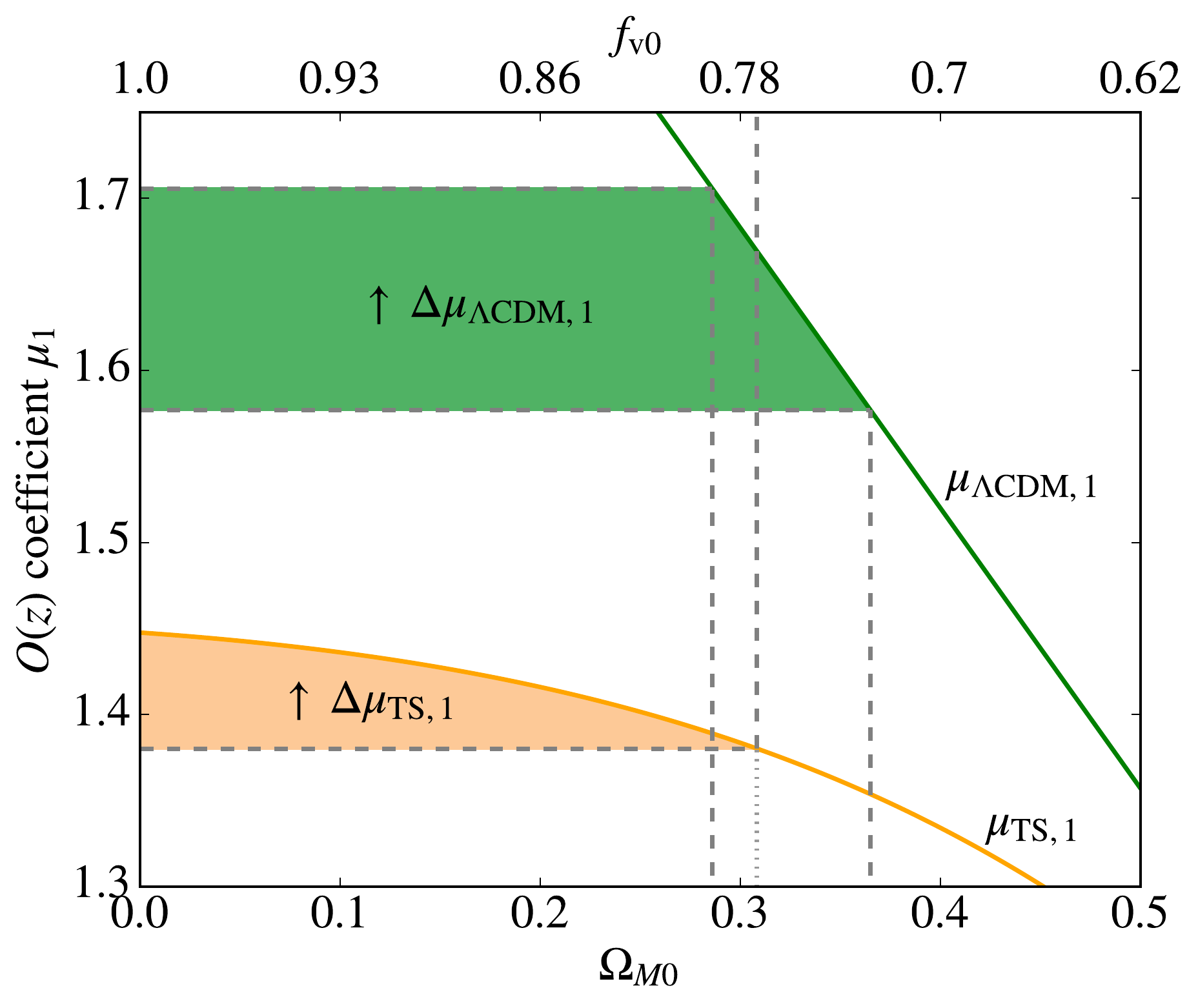}
\ \includegraphics[scale=0.35]{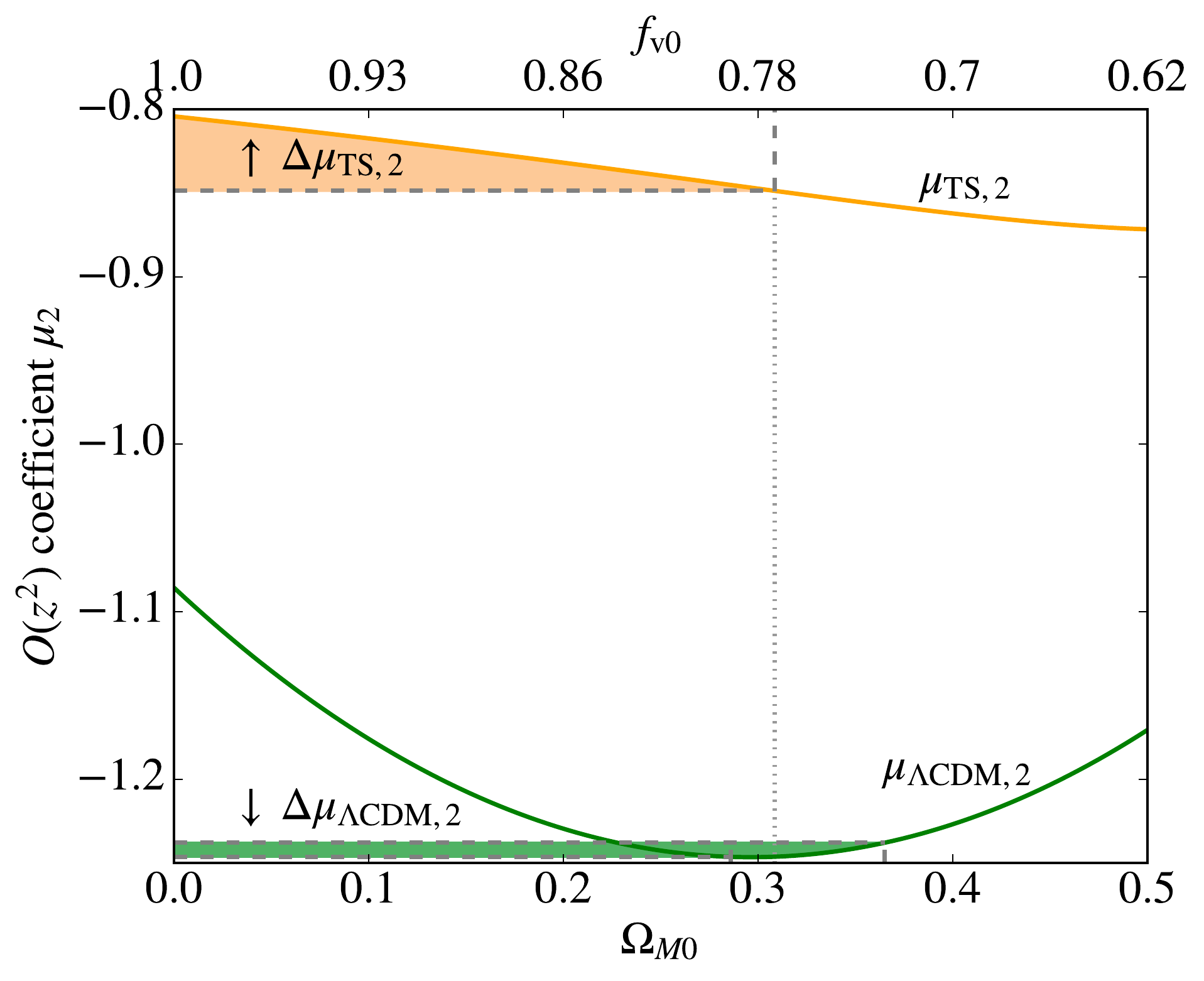}}
\centerline{\includegraphics[scale=0.35]{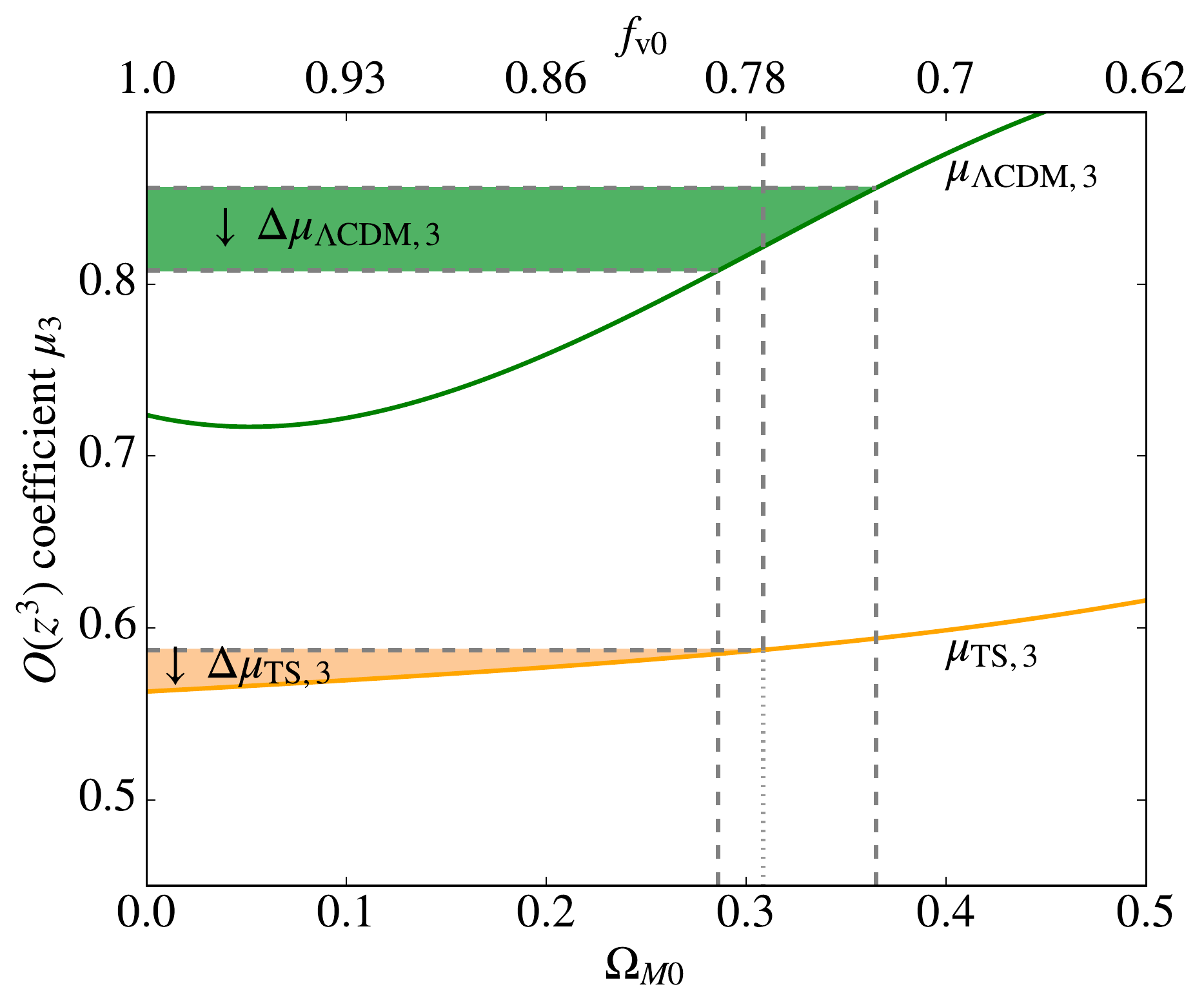}
\ \includegraphics[scale=0.35]{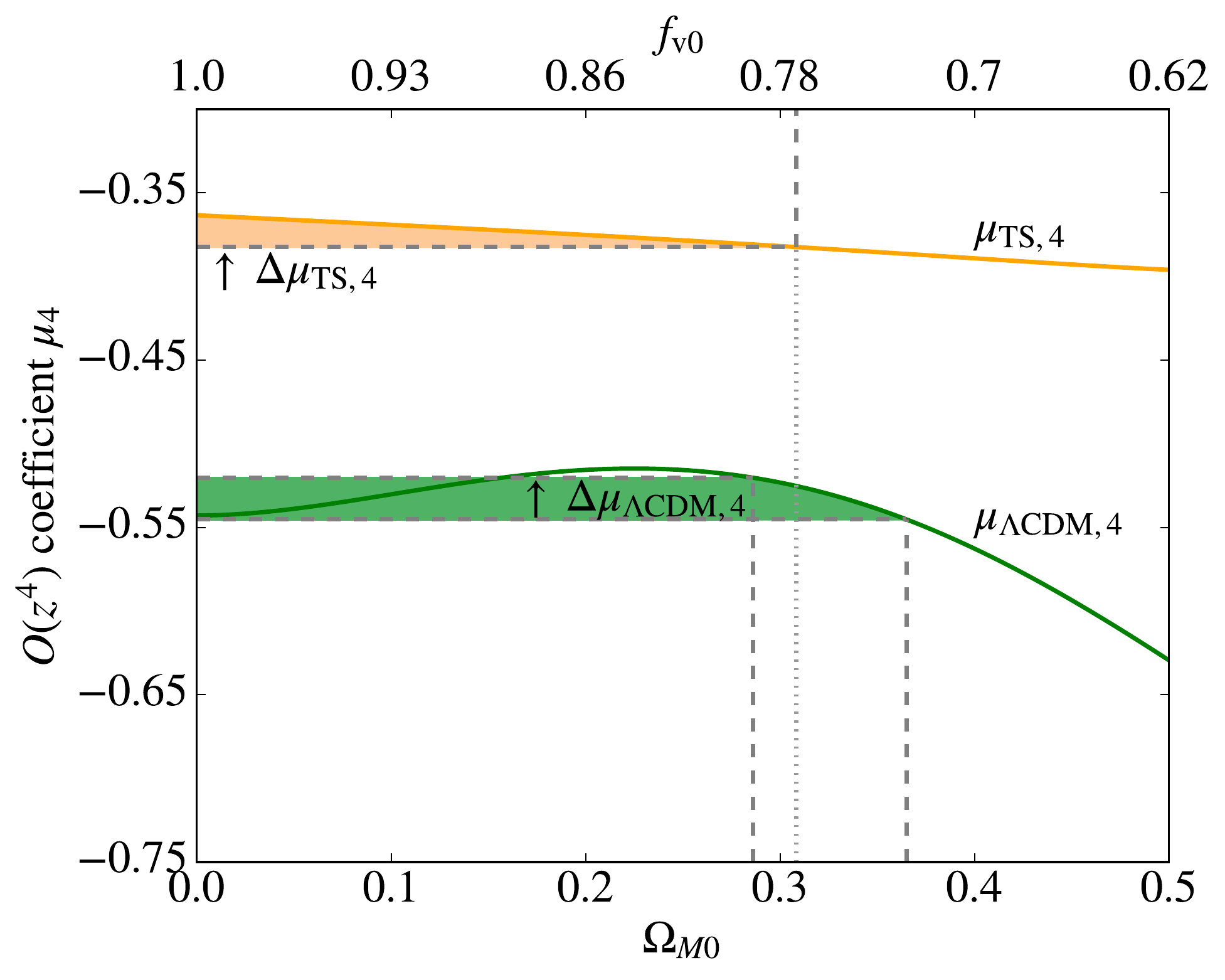}}
\caption{Coefficients in the Taylor series (\ref{muTS}),
(\ref{mulcdm}), $\mu=\mu\Z0(z)+\sum_{n=1}\mu_n z^n$,
of the spatially flat \LCDM\ and \ts\ models,
as a function of the free cosmological parameter, $\OmMn$ or $\fvn$. For \ts\
the coefficients
$\mu\Z{{\rm TS},n}$ are very slowly varying monotonic functions of $\fvn$
on the range $0.6<\fvn\le1$, whereas the coefficients
$\mu\Z{\Lambda{\rm CDM},n}$ are polynomials of order $n$. For each $n$,
$|\mu\Z{{\rm TS},n}|<|\mu\Z{\Lambda{\rm CDM},n}|$, reflecting the ``flatter''
distance modulus (cf.\ Fig.~\ref{mures}). Linear changes
of $\OmMn$ can become degenerate with empirical light-curve parameters
linear in $z$ for parameters close to the minimum of $\mu\Z{\Lambda{\rm CDM},2}$
at $\OmMn=0.296$. The change in the coefficients between NGS16 model I and
models VII/VIII is indicated.}
\label{mucoeff}
\end{figure*}

To understand the origin of such differences consider the
Taylor series expansions (\ref{muTS})--(\ref{muempty}) for the \ts,
spatially flat \LCDM\ and empty universe models, as given in
Appendix \ref{lumd}. Leading coefficients for (\ref{muTS}) and
(\ref{mulcdm}) are shown graphically in Fig.~\ref{mucoeff} as a function
of the free cosmological parameter.

All cosmologies show improvement to a global increase in $x\Z{1,0}$ with
redshift and piecewise decreases in $c\Z0$, including the empty model which
has no free parameter to adjust. However, if we incorporate linear corrections
$x_z z$ to $x\Z{1,0}$, or $c_z z$ to $c\Z0$, in the SALT2 relation, then
the difference of (\ref{salt}) and (\ref{muTS})--(\ref{muempty}) gives
a potential degeneracy between empirical parameters $x_z$ or $c_z$ and changes
in the free
cosmological parameter if the linear term in (\ref{muTS})--(\ref{muempty}) can
be changed without greatly altering the next most important O($z^2$) term.
Such a possibility is admitted by \LCDM\ but not \ts.

For \LCDM, the O($z$) term in (\ref{mulcdm}) is linear in $\OmMn$, and the
O($z^2$) term is quadratic in $\OmMn$ with a minimum at $\OmMn=
\frn8{27}=0.296$. For model V with split linear redshift laws in $c\Z0$
only, the best-fitting $\OmMn$ coincides precisely with this minimum. The decrease
in $\OmMn$ by adding a global linear $z$
dependence to $x\Z{1,0}$ is approximately the same, $\DE\OmMn=-0.01$,
in going from model V to VIII, or from model I to II. The difference in
(\ref{mulcdm}) between models I and VII/VIII,
\begin{align}&\mulcdm(0.286)-\mulcdm(0.365)=0.1287\,z- 0.0085\,{z}^{2}
\nonumber\\ &\hblank{8}-
 0.0481\,{z}^{3}+ 0.0249\,{z}^{4}+ 0.0161\,{z}^{5}-
 0.0232\,{z}^{6}+\dots
\label{dbflcdm}\end{align}
is dominated by the linear redshift changes, with negligible changes in the
O($z^2$) term.

\begin{figure*}
\centering
\centerline{{\bf(a)}\includegraphics[scale=0.43]{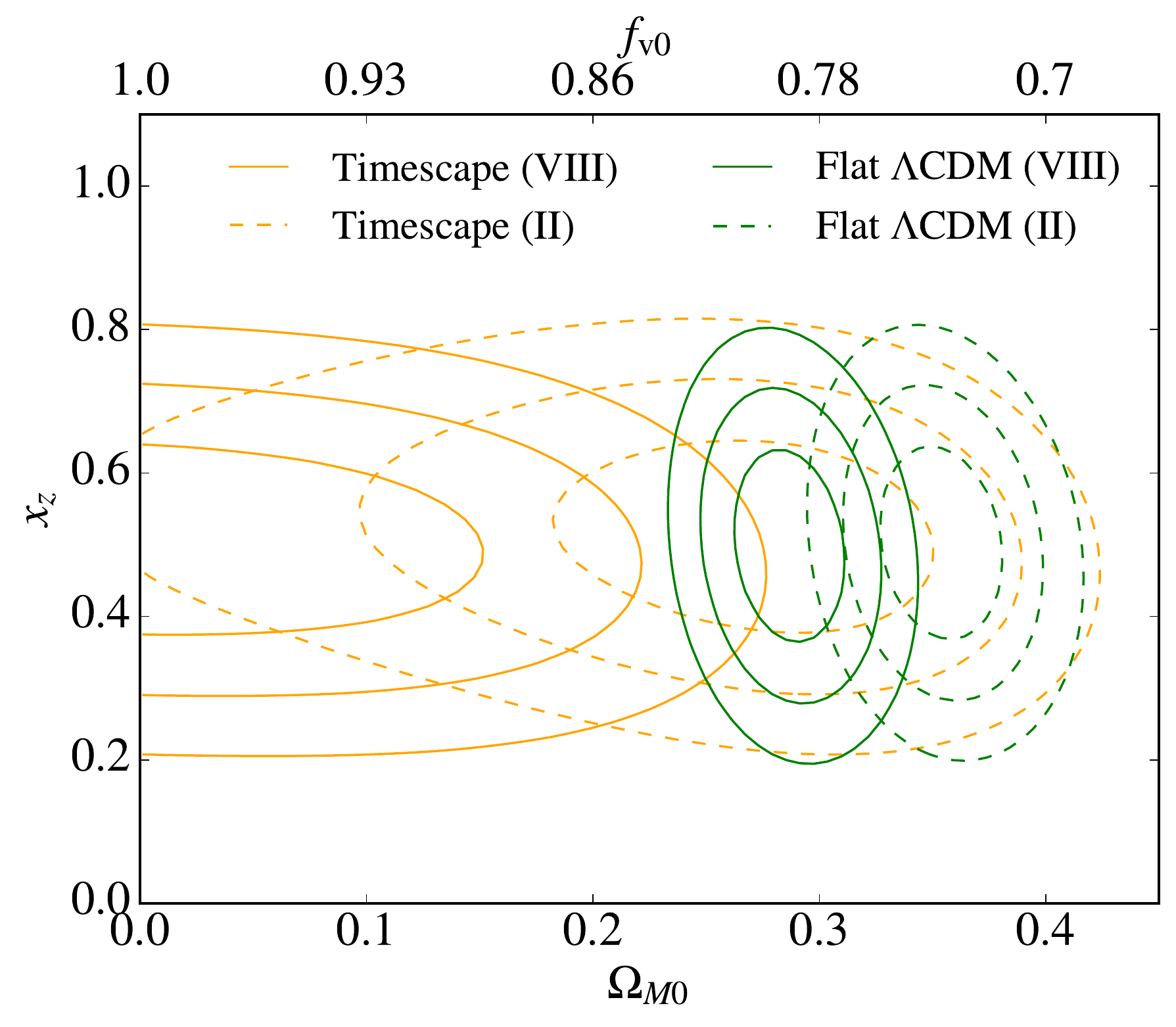}
{\bf(b)}\includegraphics[scale=0.43]{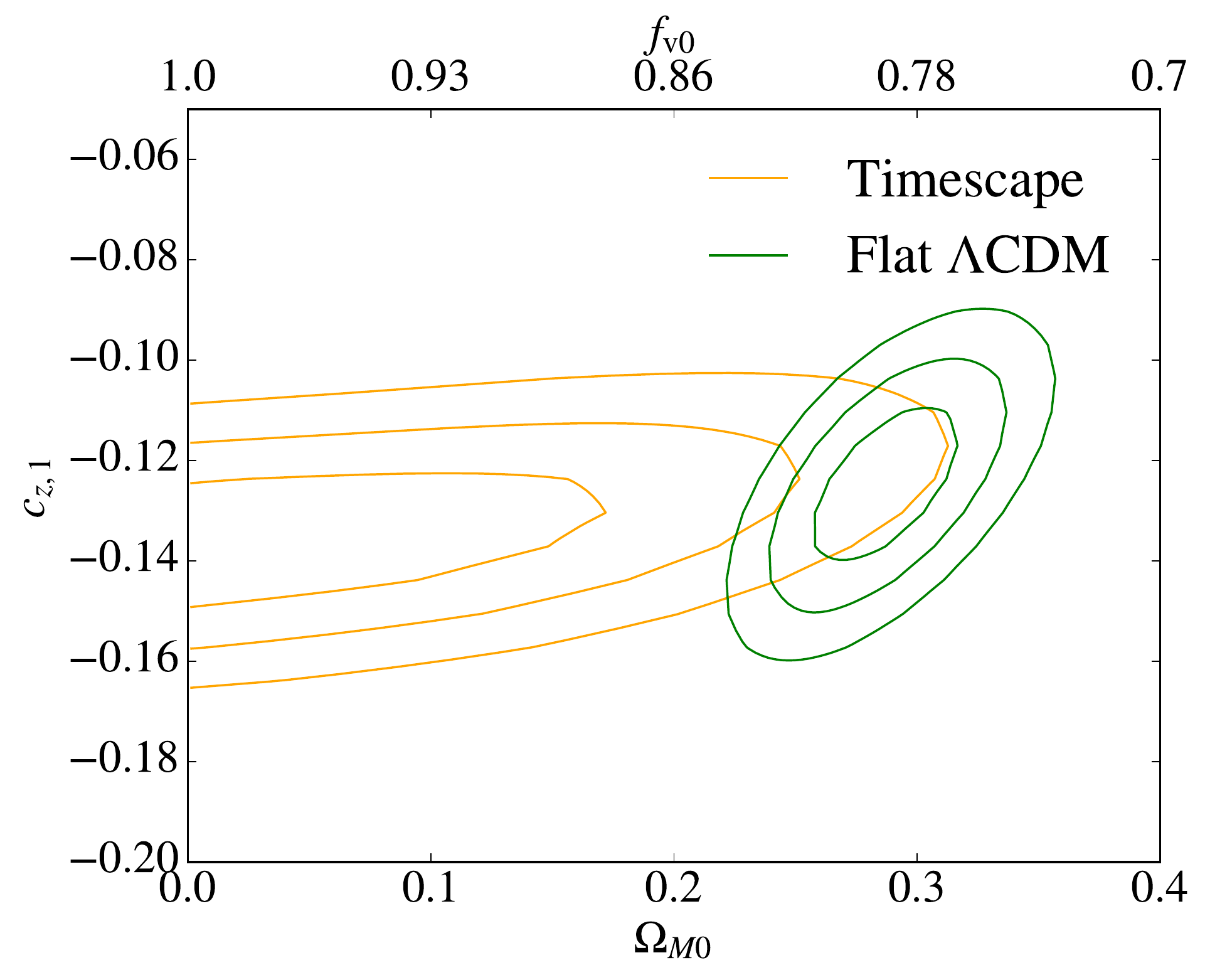}}
\centerline{{\bf(c)}\includegraphics[scale=0.43]{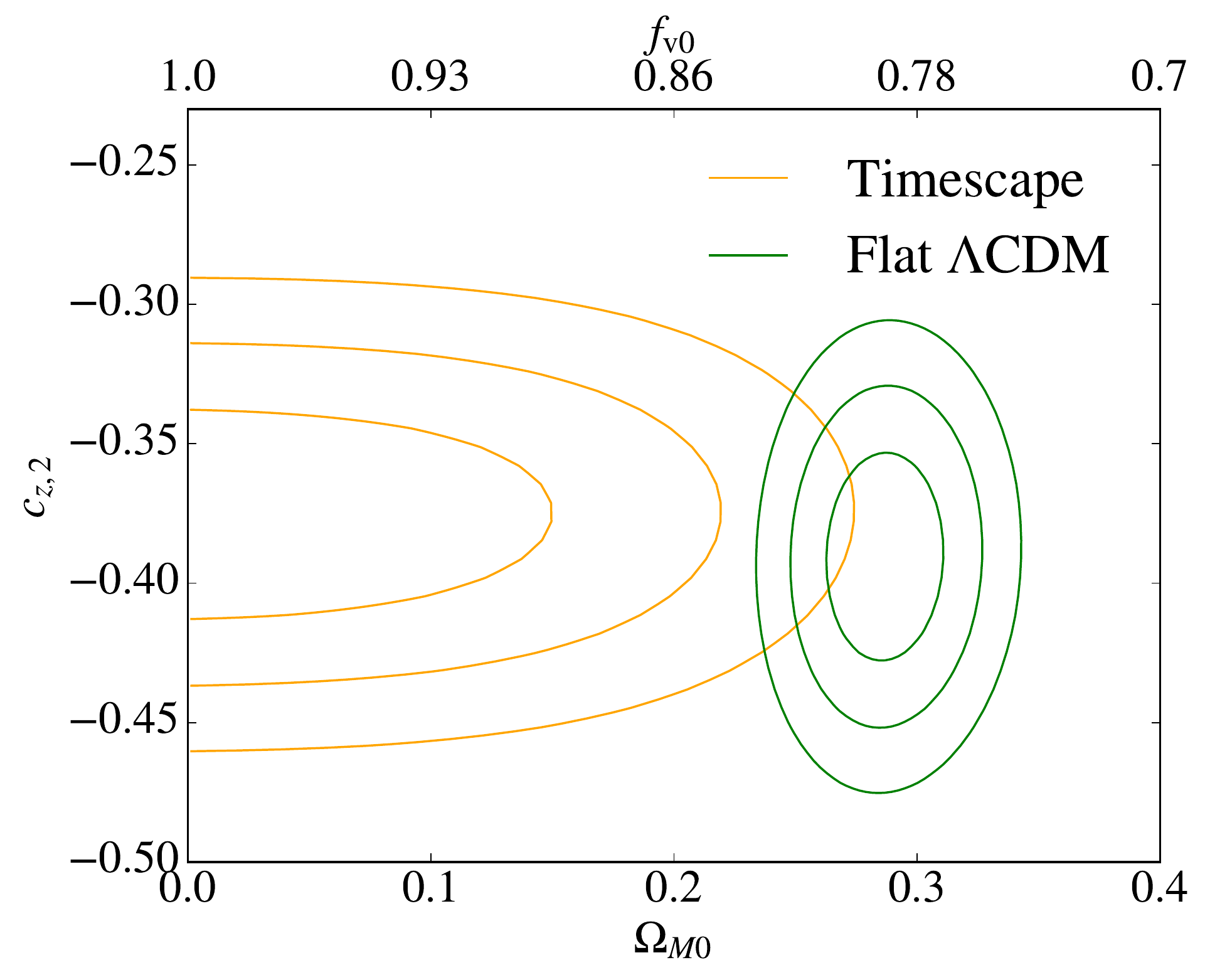}
{\bf(d)}\includegraphics[scale=0.43]{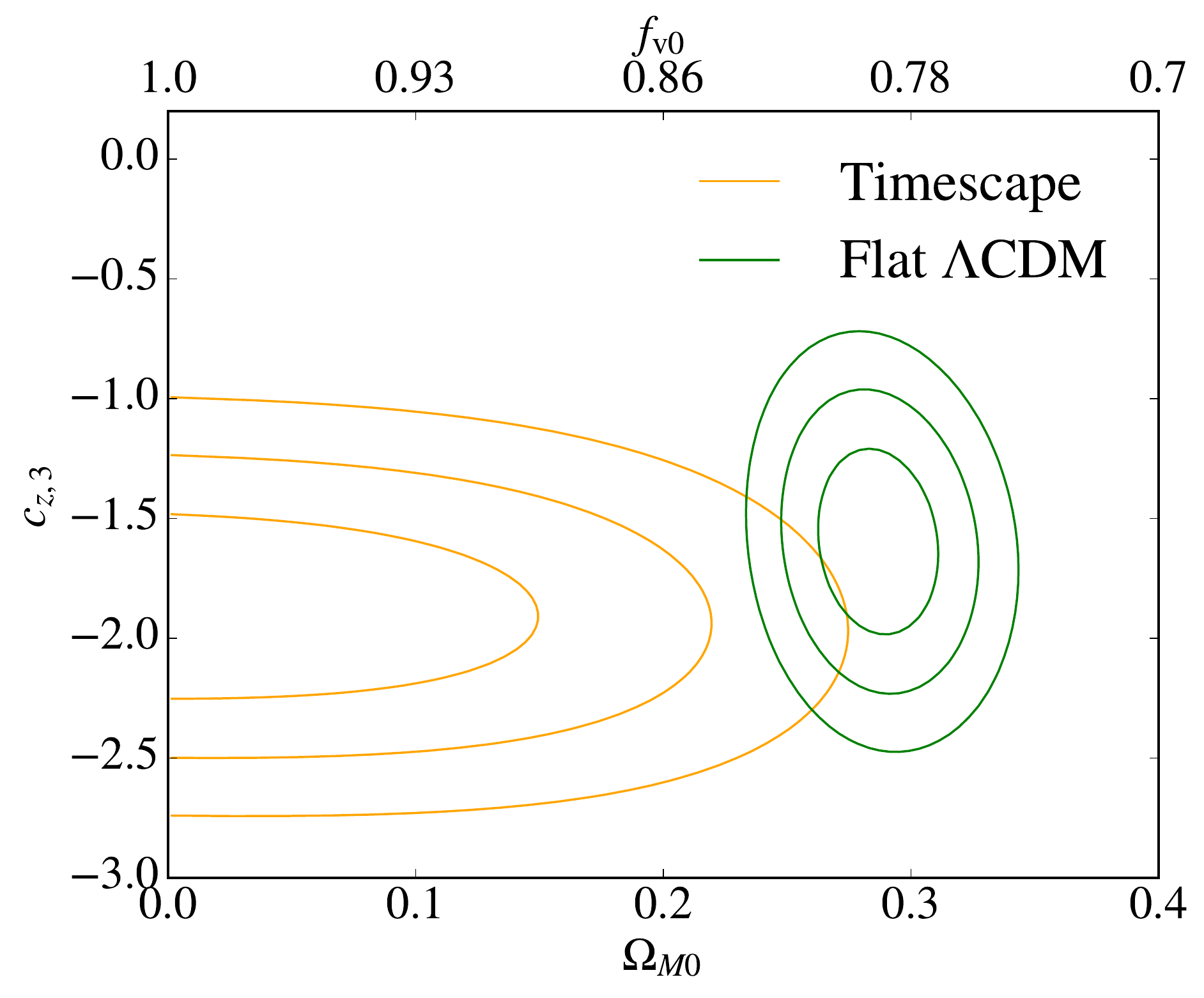}}
\caption{Likelihood function contours for model VIII with $\zmin=0.033$
projected in the planes: {\bf(a)} ($x\Z z,\OmMn$); {\bf(b)} ($c\Z{z,1},\OmMn$)
(SNLS sample, mean redshift $\ave{z}=0.636$);
{\bf(c)} ($c\Z{z,2},\OmMn$) (SDSS sample, mean redshift $\ave{z}=0.199$);
{\bf(d)} ($c\Z{z,3},\OmMn$ (low $z$ sample with $z>0.033$, mean redshift
$\ave{z}=0.0495$). 67.3\%, 95.5\%, and 99.7\%
confidence contours are shown. In panel {\bf(a)} $x\Z z$ contours for
model II are also shown to demonstrate the effect of adding the $c\Z{z,J}$
parameters. For spatially flat \LCDM\, the maximum likelihood
is driven to the vicinity of the minimum $\OmMn=\frn8{27}$ of the O($z^2$)
Taylor series term (\ref{mulcdm}). The \ts\ Taylor series (\ref{muTS})
consists of slowly varying monotonic functions of $\fvn$, and the maximum
likelihood is driven to the edge of parameter space, $\fvn\to1$.}
\label{zvar}
\end{figure*}

By contrast the terms in the Taylor series (\ref{muTS}) for \ts\ are very
slowly varying monotonic functions of $\fvn$ on the range $0.6<\fvn\le1.0$
(as shown in Fig.~\ref{mucoeff}),
so changes in $\mu\Ns{TS}$ are much more constrained. The difference in
(\ref{muTS}) between models I and VII/VIII, is
\begin{align}&\mu\Ns{TS}(1.0)-\mu\Ns{TS}(0.778)=0.0674\,z+ 0.0444\,{z}^{2}
\nonumber\\ &\hblank{8}-0.0242\,{z}^{3}
+ 0.0190\,{z}^{4}- 0.0193\,{z}^{5}+ 0.0173\,{z}^{6}+\dots
\end{align}
A large change in $\fvn$ is required make changes in $\mu\Ns{TS}$ comparable
to the \LCDM\ case, and the effect of increasing $\fvn$ increases
both the O($z$) and O($z^2$) terms. As seen in Fig.~\ref{zvar}, the likelihood
is consequently peaked along narrow ridges in the ($x\Z z$,$\fvn$) and
($c\Z{z,J}$,$\fvn$) planes, with almost constant values of $x\Z z$ and
$c\Z{z,J}$ and no upper bound on $\fvn$.

\section{Discussion}

Our study has a number of important consequences. Firstly, the \ts\ and
spatially flat \LCDM\ model luminosity distance--redshift fits to the JLA
catalogue are statistically indistinguishable using either the
approach of NGS16, or with modifications to only
the mean stretch parameter. As shown in Table~\ref{tab:var}
the Bayesian complexity, $C_b$, is lower (better) for timescape
than for \LCDM, for every choice of light-curve model.

This completely reframes a debate \citep{ngs,rh,bahamas,hldv,tldb} about
whether cosmic acceleration is marginal or not, within the confines of a FLRW
expansion history. Current supernova data does not distinguish between the
standard \LCDM\ model and the non-FLRW expansion history of the \ts\ model,
which has non-zero apparent cosmic acceleration but with a {\em marginal
amplitude}. The apparent deceleration parameter (\ref{qeff}) for the best-fitting
value of Table~\ref{tab:maxL} is $q\Z0\equiv q(\fvn)=-0.043^{+0.004}_{-0.000}$.

Within the class of FLRW models the significance of cosmic
acceleration is often assessed by comparison to the empty universe model.
That model is unphysical, since standard nucleosynthesis
and recombination can never occur in a universe with $a(t)\propto t$
regardless of its matter content.\footnote{In particular, the $R_h=ct$
model is unphysical for this reason \citep*{lbk}.} The timescape model has
positive $\ln B\Z2$ compared to the empty universe.\footnote{This is
true for the NGS16 model I and all light-curve models for which $\ln B\Z1$ shows
an improvement independent of cosmology, viz.\ models II, V, VIII.}
Nonetheless, $|\mu\Ns{TS}(z)-\mu\Ns{empty}(z)|<|\mulcdm(z)-\mu\Ns{empty}(z)|$
(c.f.\ Fig.~\ref{mures})
at late epochs, for a simple physical reason. The \ts\ model is void
dominated at $z<1$, and the expansion of individual voids is
close to an empty universe. While the \ts\ model has apparent acceleration
at late epochs, its expansion law is closer to that of an empty
universe than that of the \LCDM\ model.

The second important consequence of our study is that allowing linear changes
with redshift to the mean colour parameter, $c\Z0$, produces cosmological model
dependency. Since the redshift--distance relation of the timescape model
effectively interpolates \citep{obs,bscg} between those of spatially flat
\LCDM\ models with different values\footnote{Note that the Planck best-fitting
value $\OmMn=0.3175$ \citep{planck} is lower than the best-fitting value for the
spatially flat \LCDM\ model value $\OmMn=0.365$ from Table~\ref{tab:maxL},
consistent with the \ts\ expectation.} of $\OmMn$, particular care must be
taken with piecewise linear relations in redshift.

The improved 16 parameter model VIII (this being a better fit than the
original 21 parameter RH16 model) has positive
Bayesian evidence for \LCDM\ relative to the timescape model. However,
this is contingent on degeneracies in the likelihood
function between the free cosmological parameter and additional empirical
parameters. The RH16 parametrization allows the \LCDM\ deceleration
parameter $q\Z0=-1+\frn32\OmMn$ contained in the O($z$) term of
the Taylor series (\ref{mulcdm}) to be adjusted\footnote{For
the NGS16 model I and models VII/VIII one has best fits $q\Z0=-0.453$ and
$q\Z0=-0.571$ respectively. The respective spatially flat \LCDM\ values
quoted by \citet{rh}, namely $q\Z0=-0.412$, $-0.552$, (or $\OmMn=0.392$,
$0.299$), differ mostly on
account of our SHS cut at $\zmin=0.033$.}
near the global minimum $\OmMn=0.296$ of the O($z^2$) term in
(\ref{mulcdm}). However, the same procedure drives the \ts\ free parameter
to an unphysical limit, $\fvn\to1$.
No fundamental model underlies the empirical parametrization (\ref{RHparm}).
Variations in $c\Z0$ are most plausibly related to selection
effects, given we cannot fit them by a global law. However, selection effects
would be more correctly modelled by removing the tail of a Gaussian
distribution rather than shifting its mean linearly in redshift.

Our results show that NGS16 did not account for every possible
selection bias that remains in the JLA catalogue, consistent with
some comments of RH16. Nonetheless, NGS16 are correct to point
out the possible pitfalls in fitting SneIa data when empirical light-curve
parameters are mixed with the cosmological parameters of a single class of
cosmological models. If SneIa are to be used to distinguish cosmological
models, then systematic uncertainties and
selection biases should be corrected in as model
independent manner as possible {\em before} the data is reduced.

A related issue which remains to be explored is the extent to which
the corrections for selection biases that have already been made in the JLA
catalogue depend on the FLRW model. \citet{jla} follow a procedure of
\citet[Sec.~6.2]{m14}, who used the SNANA package to estimate selection
biases. Simulated data (using the FLRW model) is used in such estimates.
While efforts have been made to consider different dark energy equations of
state \citep{m14}, models which do not satisfy the Friedmann equation fall
outside the scope of such analyses.

Whether or not the \ts\ model is ultimately a better fit than
the standard FLRW model, it may provide a useful diagnostic tool in comparing
methods for SneIa light-curve reduction purely at the empirical level.
In particular, it has an analytic non-FLRW redshift--distance relation which is
very close to that of the \LCDM\ model, but which is considerably more
constrained in the free parameter $\fvn$ than the \LCDM\ model is in $\OmMn$.

Finally, Figs.~\ref{lcurve}(b),(c),(d) show evidence for a $\simeq100\hm$
statistical homogeneity scale which has an effect on global fits
of light-curve parameters -- most notably a 30\% shift of $c\Z0$ --
{\em independent of the cosmological model}. These systematics, which occur
at a scale relevant from independent observations \citep{h05,sdb12}, must
be explained irrespective of the cosmological model.

\section*{Acknowledgements}
We thank Chris Blake, Tamara Davis, Ahsan Nazer, David Rubin, Subir Sarkar and
Bonnie Zhang for helpful discussions and correspondence, and Thomas Buchert
for hospitality at the ENS, Lyon, France. This work was supported by
Catalyst grant CSG-UOC1603 administered by the Royal Society of New Zealand.

\vskip20pt \noindent{\em Code availability:}
The code and data used in this analysis are available at https://doi.org/10.5281/zenodo.831360\ .

\appendix
\section{Luminosity distances in the FLRW and timescape cosmologies}\label{lumd}

We compare SneIa observations to distance moduli (\ref{eq:mu})
for theoretical luminosity distances determined from the FLRW and \ts\ models.
Regardless of the matter content of the universe, the distance modulus for
any general FLRW model can be expanded as a Taylor series of derivatives of
the scale factor $a(t)$ for small redshifts, $z$. This leads to
a distance modulus \citep{Vis04}
\begin{align}
&\mu\Ns{FLRW}=25+5\log\Ns{10}\left(\CC\,z\over\Hn\,\hbox{Mpc}\right)
\nonumber\\ &+{5\over\ln10}\Bigl\{\frn12(1-q\Z0)z
+\frn1{24}\Bigl[9{q\Z0}^{2}-10q\Z0-7-4j\Z0-\Omkn\Bigr]z^2\nonumber\\
&\hblank{13}
+\frn1{24}\Bigl[s\Z0+5-10{q\Z0}^3-16{q\Z0}^2-9q\Z0+j\Z0\left(8q\Z0+7\right)
\nonumber\\ &\hblank{23}-4\Omkn\left(q\Z0+1\right)\Bigr]z^3
+\dots\Bigr\}, \label{FLRWTaylor}
\end{align}
where $\CC$ is the speed of light, and
$\Hn$, $q\Z0$, $j\Z0$, $s\Z0$ and $\Omkn$
are the present values of the Hubble, deceleration,
jerk, snap and spatial curvature parameters: $H(t)\equiv a^{-1}\pt_t a$;\
$q(t)\equiv-a^{-1}H^{-2}\pt_t^2 a$;\ $j(t)\equiv a^{-1}H^{-3}\pt_t^3 a$;\
$s(t)\equiv a^{-1}H^{-4}\pt_t^4 a$;\ $\Omk(t)\equiv-k\CC^2(Ha)^{-2}$.

The luminosity distance-redshift relation in the \LCDM\ model is given
exactly by
\begin{align}
\dL&=\frac{(1+z)\CC}{\Hn\sqrt{|\Omkn|}}\w{sinn}\left(\sqrt{|\Omkn|}\int\limits^1
_{1/(1+z)}\frac{\dd y} {\mathcal{H}(y)}\right),
\nonumber\\ &\mathcal{H}(y)\equiv\sqrt{\OmRn+\OmMn y+\Omkn y^2+\OmLn y^4}\,,
\nonumber\\ &
\w{sinn}(x)\equiv
\begin{cases} \sinh(x), &\Omkn>0\\ x,&\Omkn=0\\ \sin(x),&\Omkn<0\\
\end{cases},\label{dLlcdm}
\end{align}
\noindent
where $\OmRn$, $\OmMn$ and $\OmLn$ are respectively the present epoch values
of the radiation, non-relativistic matter and cosmological constant density
parameters, which at all epochs obey the Friedmann equation sum rule
$\OmR+\OmM+\Omk+\OmL=1$. Since
$\OmRn=4.15\times10^{-5}h^{-2}$, the radiation term can be neglected on the
distance scales of supernovae. Furthermore, for FLRW models $\Omkn$ is
constrained to be close to zero by the angular diameter distance of the
sound horizon in the CMB. Thus we
will restrict attention to the spatially flat \LCDM\ model, with two effective
free parameters, $\Hn$ and $\OmMn\simeq1-\OmLn$. We use eq.\
(\ref{dLlcdm}) with $\Omkn=0$, $\OmRn=0$, for computations but note that in
the Taylor series (\ref{FLRWTaylor}),
$q\Z0=-1+\frn32\OmMn$, $j\Z0=1$,
$s\Z0=1-\frn92\OmMn$.

We also consider the FLRW model with linear
expansion $a(t)\propto t$. This solution is obtained by setting
$\Omkn=1$, $\OmRn=\OmMn=\OmLn=0$ in (\ref{dLlcdm})
or $\Omkn=1$, $q\Z0=j\Z0=s\Z0=\dots=0$ in (\ref{FLRWTaylor}).
Following NGS16 we denote this the {\em empty universe}, but note
any matter content is admissible as long as the
luminosity distance is exactly $\dL=\CC\,z(1+\frn12z)/\Hn$.\medskip

The \ts\ model \citep{clocks,sol,obs,dnw}, does not evolve by the Friedmann
equation, and its distance modulus does not yield a Taylor series
that coincides with (\ref{FLRWTaylor}) beyond the leading Hubble term.
Instead observables are determined by conformally
matching radial null geodesics of the regional ``finite infinity''
geometry of observers in gravitationally bound systems to a statistical
geometry determined by fitting a spherically symmetric metric
to a solution \citep{sol,obs,dnw} of the Buchert equations \citep{be1,be2}.

For the purpose of supernova distance analysis, the radiation density
parameter (though somewhat differently calibrated to the CMB \citep{dnw}) is
negligible at the present epoch. To an accuracy of 0.3\% the expansion
history at late epochs is then given analytically \citep{sol,obs}.
The ``dressed'' luminosity distance, $\dL$, and angular
diameter distance, $\dA$, are given by
\begin{align}
\dL=&(1+z)^2\dA,\label{dATS}\\
\dA=&\CC\,t^{2/3}\int_t^{t\X0}
{2\,\dd \tb\over(2+\fv(\tb))(\tb)^{2/3}}\nonumber
\\ =&\CC\,{t^{2/3}(\FF(t\Z0)-\FF(t))},
\label{dLTS}\\
\FF(t)&\equiv2t^{1/3}+{b^{1/3}\over6}\ln\left((t^{1/3}+b^{1/3})^2\over
t^{2/3}-b^{1/3}t^{1/3}+b^{2/3}\right)\nonumber\\
&\hblank{20}+{b^{1/3}\over\sqrt{3}}\tan^{-1}\left(2t^{1/3}-b^{1/3}\over
\sqrt{3}\,b^{1/3}\right),\label{FF}
\end{align}
where the volume-average time parameter, $t$, is defined implicitly in terms
of the redshift by
\beq
z+1={(2+\fv)\fv^{1/3}\over3\fvn^{1/3}\Hb t}
={2^{4/3}t^{1/3}(t+b)\over\fvn^{1/3}\Hb t(2t+3b)^{4/3}}\,,
\label{redshift}\eeq
$b\equiv2(1-\fvn)(2+\fvn)/(9\fvn\Hb)$,
$\fvn$ is the present epoch value
of the void volume fraction,
\beq\fv(t)={3\fvn\Hb t\over3\fvn\Hb t+(1-\fvn)(2+\fvn)}\,,
\eeq
and $\Hb$ is the ``bare Hubble
constant'' related to the observed
Hubble constant by $\Hb=2(2+\fvn)\Hn/(4{\fvn}^2+\fvn+4)$. The parameter $t$
is related to the time parameter, $\tau$, measured by typical observers in
bound structures by
\beq
\tau=\frn23 t+{2(1-\fvn)(2+\fvn)\over27\fvn\Hb}\ln\left(1+{9\fvn\Hb t
\over2(1-\fvn)(2+\fvn)}\right)\,. \label{tsol}
\eeq

The effective dressed scale factor $a(\tau)$ is given by
\beq a\equiv\gb^{-1}\ab,\eeq where $\ab$ is the bare or volume-average
scale factor and $\gb$ is the phenomenological lapse function. These
have simple analytic forms in terms of the volume-average time parameter, $t$,
namely
\beq
\ab={\ab\Z0\bigl(3\Hb t\bigr)^{2/3}\over2+\fvn}\left[3\fvn\Hb t+
(1-\fvn)(2+\fvn)\right]^{1/3}
\eeq
and
\beq
\gb=\frn12(2+\fv)={3(t+b)\over(2t+3b)}
\eeq
respectively \citep{sol,obs}. The bare Hubble parameter,
$\bH\equiv\pt_t\ab/\ab$, and dressed Hubble parameter, $H\equiv\pt_\tau a/a$,
are given respectively by
$\bH=(2+\fv)/(3t)$ and $H=\left(4\fv^2+\fv+4\right)\bH/[2(2+\fv)]$.
The bare deceleration parameter, $\mean q\equiv-\ab^{-1}\bH{}^{-2}\pt_t^2\ab$,
is always positive. However, on account of the different time parameters
(\ref{tsol}) the dressed deceleration parameter inferred by observers
in bound systems,
$q\equiv-a^{-1}H^{-2}\pt_\tau^2 a$, may change sign from positive to negative,
indicating apparent acceleration. The dressed deceleration parameter is
given by
\beq
q={-\fvf(8\fv^3+39\fv^2-12\fv-8)\over\left(4+\fv+4\fv^2\right)^2}\,.
\label{qeff}\eeq
The onset of apparent acceleration is determined by a root of the cubic
in $\fv$ in the numerator of (\ref{qeff}), and begins when $\fv\simeq0.587$.

\newpage\begin{widetext}
One may substitute (\ref{dATS})--(\ref{FF}) in (\ref{eq:mu}) and then
invert (\ref{redshift}) as a series in $z$ using an algebraic computing
package to arrive at a low redshift Taylor series for the distance modulus,
$\mu\Ns{TS}$, equivalent to (\ref{FLRWTaylor}) for the FLRW model.
The first terms are given below,
along with equivalent expressions for the spatially flat \LCDM\ and empty
universe models as determined from (\ref{FLRWTaylor}):
\begin{align}
&\mu\Ns{TS}=\mu\Z0(z)+{5\over\ln10}\left\{
\left[\frac {24\,\fvn^{4}-23\,\fvn^{3}+99\,\fvn^{
2}+8}{2\left( 4\,\fvn^{2}+\fvn+4 \right) ^{2}}\right]z\right.,\nonumber\\
&\hbox to 30pt{\hfil}-\left.\left[
{\frac {1984\,\fvn^{8}-4352\,\fvn^{7}+16515\fvn^{6}+14770\,\fvn^{5}+7819\,\fvn^{4}-11328\fvn^{3}+32080\,\fvn^{2}-128\,\fvn+960}{24\left( 4\fvn^{2}+\fvn+4 \right) ^{4}}}\right]z^2
+\dots\right\},\label{muTS}\\
&\mulcdm=\mu\Z0(z)+\frn5{\ln10}\left\{(1-\frn34\OmMn)z
-\left[\frn12+\frn12\OmMn-\frn{27}{32}\OmMn^2\right]z^2
+\left[\frn13-\frn18\OmMn+\frn{21}{16}\OmMn^2-\frn{45}{32}\OmMn^3\right]z^3
+\dots\right\}\label{mulcdm}\\
&\mu\Ns{empty}=\mu\Z0(z)+\frn5{\ln10}\left\{\frn12z-\frn18z^2+\frn1{24}z^3
+\dots\right\}.\label{muempty}
\end{align}
Here the term $\mu\Z0(z)\equiv25+5\log\Ns{10}[\CC z/(\Hn\w{Mpc})]
=25+5\log\Ns{10}(2997.9\,h^{-1})+5\log\Ns{10}z$
 is common to all models, the Hubble constant being
$\Hn=100\,h\kmsMpc$.
\end{widetext}

\section{Implementation of the SALT2 method}\label{salt2}

The SALT relation (\ref{salt}) refers to the actual emitter (em) and
observer (obs),
but the luminosity distance relations (\ref{dLlcdm}) and (\ref{dLTS})
refer to ideal observers who determine an isotropic distance--redshift
relation. Consequently, the theoretical relations (\ref{dLlcdm}) and
(\ref{dLTS}) must be transformed to the frame involving the actually measured
redshift $\zh=(\la\ns{obs}-\la\ns{em})/\la\ns{em}$ before the
SALT relation is applied. The luminosity distance entering (\ref{eq:mu}) is
then
\beq
\hat\dL(\zh)=\frac{1 + \zh}{1 + z} \dL(z)=(1+\zh)D(z),
\eeq
where for each cosmological model, $D(z)=\dL/(1+z)=(1+z)\dA$ is the
(effective) comoving distance, and
\beq
1+\zh=(1 + z)(1+z^{\rm pec}\ns{obs})(1+z^\phi\ns{obs})
(1+z^{\rm pec}\ns{em})(1+z^\phi\ns{em})\label{redsplit}
\eeq
gives the measured redshift, $\zh$, in terms of the cosmological redshift, $z$,
the local Doppler redshifts of observer, $z^{\rm pec}\ns{obs}$, and emitter,
$z^{\rm pec}\ns{em}$, and gravitational redshifts at the two locations,
$z^\phi\ns{obs}$ and $z^\phi\ns{em}$.

For our observations, $\zh$, is the heliocentric redshift as the Earth's
annual motion is averaged to the rest frame of the Sun. In the standard
cosmology gravitational potential effects are assumed to be small, and the
only relevant terms on the r.h.s.\ of (\ref{redsplit}) are assumed to be
local boosts of order $v/c\goesas10^{-3}$. This leads to 0.1\% corrections to
the luminosity distance which are often neglected. However, as noted by
\citet{cd} differences of 0.1\% in $\dL$ lead to
order 1\% corrections to cosmological parameters, which we have confirmed
in our analysis.

In the \ts\ model, as in any inhomogeneous cosmology, expansion below
the $\goesas100\hm$ SHS will generally differ from that of a global average
geometry plus local boosts. Equivalently, very slowly varying time-dependent
gravitational potentials also make a contribution to (\ref{redsplit}). Such
terms encode non-kinematic differential expansion \citep*{bnw} from inhomogeneities
below the SHS. Spatial variations in the term $z^\phi\ns{em}$ may have
significant consequences for interpreting the local ``peculiar velocity
field'' for sources within $\lsim100\hm$ of our location \citep{rest,mw} but any
net anisotropy on SneIa redshifts on larger scales should only make a small
correction to the standard boost between the heliocentric and CMB frames.
Indeed, it could be a source for a small systematic redshift uncertainty of the
type considered by \citet{cd}. However, we do not investigate
that possibility in the present paper as the RH16 empirical light-curve models
we study are already very complex.
The peculiar velocity and gravitational potential terms in (\ref{redsplit})
that we are unable to determine will be assumed to contribute to statistical
uncertainties in measured redshifts only.

We therefore compute cosmological luminosity distances is the CMB rest frame,
exclude data below the SHS, and study the effect of different choices
for this cutoff.\footnote{Since we do not constrain $\Hn$, we do not
specifically investigate the relationship between light-curve parameters
and determinations of the local Hubble constant, which have been discussed
in the past \citep{JRK,c07,h09,sw,rest,mw}. In the \ts\ model higher average
values of $\Hn$ are expected
below the SHS.} Furthermore, we apply the SALT2 relation in the heliocentric
frame using the values tabulated in the JLA catalogue \citep{jla}, and calculate
the corresponding CMB rest frame redshifts ourselves.\footnote{We use the
NASA/IPAC Extragalactic Database standard, $371\kms$ in the direction
$(\ell,b)=(264.14^\circ,48.26^\circ)$ \citep{f96}.}

We {\em do not} use the JLA tabulated CMB frame redshifts \citep{jla} since in
addition to our $z^{\rm pec}\ns{obs}$ correction, these values also already include a
correction, $z^{\rm pec}\ns{em}$, for the peculiar velocity
field \citep{hslb4,c11} of galaxies up to $z=0.071$, implicitly assuming
the FLRW model.

\section{Model comparison}\label{stat}
\subsection{Frequentist approach} \label{A}

We are interested in the dependence of the likelihood (\ref{eq:likefin}) on
the model parameters, $\Theta$. We write $\mathcal{L}(\Theta)
\equiv \mathcal{L}( \mathrm{Data} | \Theta, \Mmod )$. We are interested
in a subset of parameter-space $\Theta\Z1\subset\Theta$, for which
we construct a profile likelihood 
$\mathcal{L}_p (\Theta\Z1) \equiv \max_{\Theta_2}[\mathcal{L}
(\Theta)]$, where maximization is over the nuisance parameters
$\Theta_2 = \Theta \setminus \Theta\Z1$.
In our case $\Theta\Z1$ is usually the free parameter(s) of the
cosmological model, and $\Theta_2$ the intrinsic supernovae parameters
and the empirical parameters, $\alpha$, $\beta$, of the light-curve fitter.

Confidence regions for the parameters of interest are estimated from
the coverage probability $p\ns{cov}$ of a region in the
$k$-dimensional slice of parameter space, $k\equiv\mathop{\rm dim}\Theta
\Z1$, given asymptotically by the integral
\begin{equation}\label{eq:normalnb}
p\ns{cov} (\mathrm{region}) = \int_{0}^{-2 \ln\left(\mathcal{L}_p(\Theta\X1)
/\mathcal{L}_{\rm max}\right)}
 f_{\chi^2} (\Theta\Z1' , k )\dd\Theta\Z1'
\end{equation}
where $\mathcal{L}_{\rm max}\equiv\max_{\Theta\X1}[
\mathcal{L}_p (\Theta\Z1)]$, and
$f_{\chi^2}(x,k)$ is the probability density function of
a $\chi^2$ distributed variable with $k$ degrees of freedom.
Having constructed confidence intervals from (\ref{eq:normalnb}), one
can compare nested models.

Since we wish to compare independent non-nested models,\footnote{We note that
only models II, IV, VI are extensions of the 9 parameter base model, i.e.,
model I is nested in II, IV and VI, while II and IV are nested
models of VI. Model V is nested in model VIII.} we need to go beyond
the procedure of (\ref{eq:normalnb}). The Akaike Information Criterion (AIC)
 \citep{aic} and Bayesian Information Criterion (BIC) \citep{bic} are two widely
used measures of the relative information loss for non-nested models, given
respectively by
\begin{align}
{\rm AIC}&=2k-2\ln\left(\mathcal{L}_{\rm max}\right)\label{AICs}\\
{\rm BIC}&=k\ln N-2\ln\left(\mathcal{L}_{\rm max}\right)\label{BICs}
\end{align}
where $k$ is the number of independent model parameters, and $N$ the number of
data points fit. The AIC estimate of relative probability of
minimal information loss for two models is $P_{AIC}\equiv\exp\left[-\frn12
\left({\rm AIC}\Z1-{\rm AIC}\Z2\right)\right]$, and similarly for BIC. The BIC
gives a greater penalty for introducing additional parameters than
AIC if $N\ge8$. Differences of at least 2, 6 and 10 are
considered to be respectively `positive', `strong' and `very strong' evidence
 \citep{kr} for the model with the lower IC value. Both tests reduce to a maximum
likelihood comparison when $k_1 = k_2$, as is the case for the \ts\ and
spatially flat \LCDM\ models.

\subsection{Bayesian approach} \label{B}
The frequentist methods place emphasis on the maximum likelihood, which
is of limited use. We therefore perform a fully Bayesian analysis of the JLA
data set to determine the relative statistical support for each cosmological
model, as well as for the introduction of additional redshift dependent light
curve parameters.

Given data, $D$, and a model, $\Mmod$, determined by a set of $n$ parameters
$\Theta=(\theta_1,\theta_2,\ldots,\theta_n)$, by Bayes theorem
the {\em posterior} probability distribution, $p(\Theta|D,M)$, is given by
\beq\label{eq:bayes}
p(\Theta|D,\Mmod) = \frac{\mathcal{L}(\Theta) \pi(\Theta|\Mmod)}{p(D|\Mmod)},
\eeq
where $\mathcal{L}(\Theta)\equiv p(D|\Theta,\Mmod)$ is the likelihood,
$\pi(\Theta|\Mmod)$ is the {\em prior} distribution and $p(D|\Mmod)$ is the
{\em Bayesian evidence}.
The prior represents a subjective initial state of belief in the parameters
based on external information or previous experiments, while
the evidence is a normalization constant,
\beq\label{eq:evidence}
E\equiv p(D|\Mmod)=\int \mathcal{L}(\Theta) \pi(\Theta|\Mmod)\,\dd\Theta,
\eeq
to ensure the posterior is a probability distribution.
It is independent of parameters and as such does not play a role in parameter
estimation, but becomes important for model comparison.

Given two models, $\Mmod_1$ and $\Mmod_2$, for the same data $D$, the Bayes
factor \citep{kr}
\beq
B\equiv\frac{E_1}{E_2}=\frac{p(D|\Mmod_1)}{p(D|\Mmod_2)},
\eeq
gives a measure for which model is more probable in view of the data. The
Bayes factor implicitly applies the principle of Occam's razor\footnote{The
AIC and BIC statistics also include a penalty using simple approximations
to Bayesian methods which derive from different assumptions about the priors.
The factor of two difference in the IC evidence scale \citep{kr} reflects the
factor of 2 multiplying $\ln\left(\mathcal{L}_{\rm max}\right)$
in the definitions (\ref{AICs}), (\ref{BICs}).} with a
penalty for adding extra parameters.
This makes model selection natural in the Bayesian framework.
Values of $B>1$ indicate preference for model 1, $B<1$ for model 2.
On a standard scale, evidence with $|\ln B|<1$ is `not worth more than a bare
mention' \citep{kr} or `inconclusive' \citep{t07}, while $1\leq|\ln B|< 3$,
$3\leq|\ln B|<5$ and $|\ln B|\ge5$ indicate `positive', `strong' and
`very strong' evidences respectively \citep{kr}.

In the Bayesian approach the nuisance parameters are {\em marginalized} over,
i.e., integrated out from the posterior $p(\Theta|D,\Mmod)$. E.g.,
the {\em marginal posterior} of $\theta_1$ is obtained from the
$n$-dimensional posterior by
\beq
p(\theta_1|D,\Mmod) = \int p(\theta|D,\Mmod)\,\dd\theta_2\,\dd\theta_3\ldots\dd\theta_n,
\eeq
and from this 1-dimensional distribution parameter inferences can be made.
The posterior mean value is given by
\beq
\bar{\theta}_1 = \int \theta_1\, p(\theta_1|D,\Mmod)\,\dd\theta_1,
\eeq
and more generally
\beq
\bar{f} = \int f(\theta_1)\, p(\theta_1|D,\Mmod)\,\dd\theta_1,
\eeq
for some parameter dependent quantity $f$.
Credible intervals, or uncertainties in parameters, can also be obtained from
the marginal posterior.
E.g., a $68\%$ equal-tailed credible
interval is defined in such a way that $(1-0.68)/2=0.16$ of the probability
lies on either side of the interval.

In cases where the Bayes factor is close to unity giving no clear preference for
either model, the {\em Bayesian complexity} \citep{spiegelhalter2002} can provide
a secondary measure to the model selection process.
It is defined as
\beq\label{eq:bayesComplex}
C_b \equiv -2\left(D\Ns{KL}(p,\pi)-\widehat{D\Ns{KL}}\right),
\eeq
where
\beq
D\Ns{KL}(p,\pi)\equiv\int p(\Theta|D,\Mmod)\ln\left[\frac{p(\Theta|D,\Mmod)}{\pi(\Theta|\Mmod)}\right]\dd\Theta,
\eeq
is the {\em Kullback-Leibler divergence} measuring the information gain
of the inference, and $\widehat{D\Ns{KL}}$ is a point estimator
evaluated at the posterior mean $\bar{\Theta}$ measuring the expected information gain:
\beq
\widehat{D\Ns{KL}}\equiv p(\bar\Theta|D,\Mmod)\ln\left[\frac{p(\bar\Theta|D,\Mmod)}{\pi(\bar\Theta|\Mmod)}\right]
=\ln\mathcal{L}(\bar\Theta)-\ln p(D|\Mmod),
\eeq
where we have used Bayes theorem in the second equality.
As the data may not be able to constrain all parameters, the Bayesian complexity
determines the effective number of parameters supported by the data.
Thus for models with $|\ln B|<1$, the model with the lower $C_b$ indicates the simpler
model and is therefore preferred.
By defining the {\em effective chi-squared} $\chi^2(\Theta)\equiv-2\ln\mathcal{L}$ and
invoking Bayes theorem \eqref{eq:bayes}, we can rewrite \eqref{eq:bayesComplex} as
\beq
C_b=\overline{\chi^2(\Theta)}-\chi^2(\bar{\Theta}),
\eeq
with $\overline{\chi^2}$ being the posterior mean of $\chi^2$.

\section{Cosmological model priors}\label{prior}

We construct priors for the \ts\ model \citep{clocks,sol,obs,dnw}
based on CMB and BAO observations, to the best of our knowledge. We will
also construct equivalent priors for the \LCDM\ model based on the same
assumptions. The resulting priors are wider than in conventionally assumed,
but do not unfairly weight a Bayesian comparison by integrating the \LCDM\
model likelihood function over a narrow parameter range as compared to the \ts\
case.\footnote{If we were to use conventional narrower priors for \LCDM\
then the \ts\ model is either unfairly advantaged or disadvantaged, depending
on whether the maximum likelihood lies within the range of the narrower prior
or not. For the NGS16 model, for example, this is not the case for the
spatially flat \LCDM\ model, and the \ts\ model is unfairly advantaged.
For model VIII the situation is reversed.}

\subsection{CMB acoustic scale constraint}
In the case of the CMB, a cosmology independent analysis of the angular
scale and heights of the acoustic peaks has been undertaken by
\citet{asas} from the Planck data. We
use the information resulting from the angular scale of the acoustic peaks
alone. The angular scale depends on the angular diameter distance of the
sound horizon alone, which is constrained in the \ts\ model. By contrast,
the relative peak heights are related to the baryon--to--photon ratio,
$\etBg$, and the spectral index, $n_s$, which are parameters with the
largest systematic uncertainties in the \ts\ case.

A non-parametric fit of the acoustic scale alone gives $286\le\ell\Ns{A}
\le305$ at 95\% confidence \citep{asas}. Our CMB prior is then determined
by demanding that the angular diameter distance of the sound horizon
at decoupling matches the corresponding angular scale $\theta\Ns{A}=\pi/
\ell\Ns{A}$; i.e., $0.01030\le\th\Ns{A}\le0.01098$. In earlier
work \citep{lnw,sw,dnw}, given that non-parametric fits had not been performed,
we had demanded a match to the FLRW parametric estimate of the acoustic scale
$\th\Ns{A}=0.01041$ to within 2\%, 4\% or 6\%. The non-parametric
fit represents a considerable improvement, particularly since the
FLRW model value is not in the mid-range of the non-parametric
95\% confidence interval.

To constrain the angular diameter distance of the sound horizon $d\ns{A\,dec}
={\bar D}_s(z\ns{dec})/\th\Ns{A}$ in the \ts\ model, we determine the redshift
of decoupling, $z\ns{dec}$, and the comoving distance of the sound horizon
$\bar D_s$ at that epoch \citep{dnw,nw}, which require the baryon--to--photon
ratio to be specified.
In the FLRW model this ratio is very tightly constrained by the ratio of
CMB peak heights, as first measured by WMAP \citep{wmap}. However, to achieve a
similarly precise constraint in the \ts\ model we would need to include
backreaction in the primordial plasma \citep{nw}, which is beyond the
scope of current investigations. In previous work \citep{dnw,lnw,sw} we used
a range of pre-WMAP baryon--to--photon ratios \citep{bbn2},
$4.6<10^{10}\etBg<5.6$, for
which all light element abundance measurements are within $2\si$, i.e.,
with no primordial lithium abundance anomaly. In the present analysis, we
wish to use the same priors on $\etBg$ for both the \ts\ and \LCDM\ models,
and thus need to include the standard model value
$\etBg=2.736\times10^{-8}\OmMn h^2=(6.08\pm0.07)\times10^{-10}$ for which the
primordial lithium abundance is problematic. We
therefore adopt the more conservative pre-WMAP range given by
\citet{bbn1}, namely $4.2<10^{10}\etBg<6.3$.

\begin{figure*}
\centering
\centerline{\includegraphics[scale=0.43]{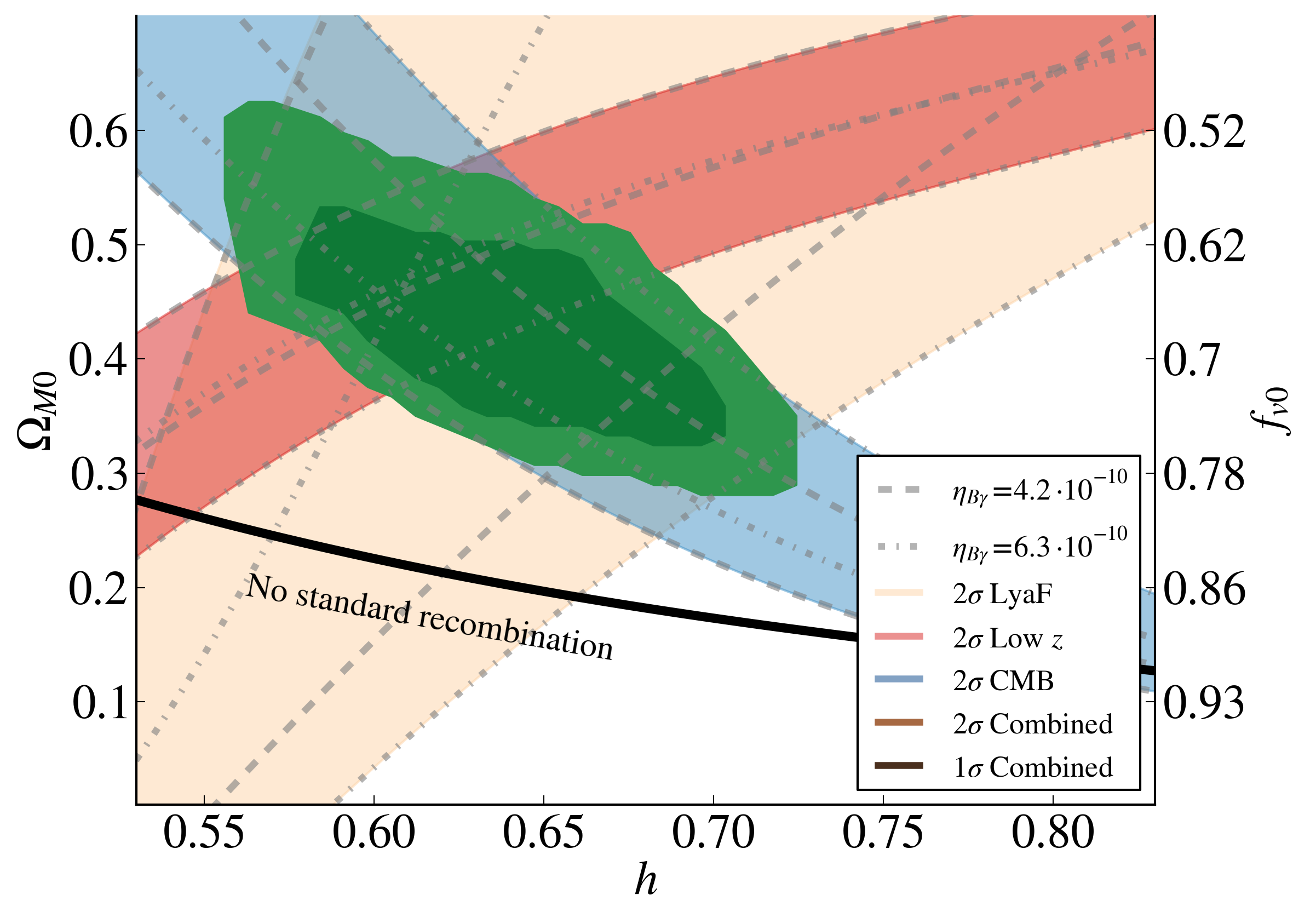}
\includegraphics[scale=0.43]{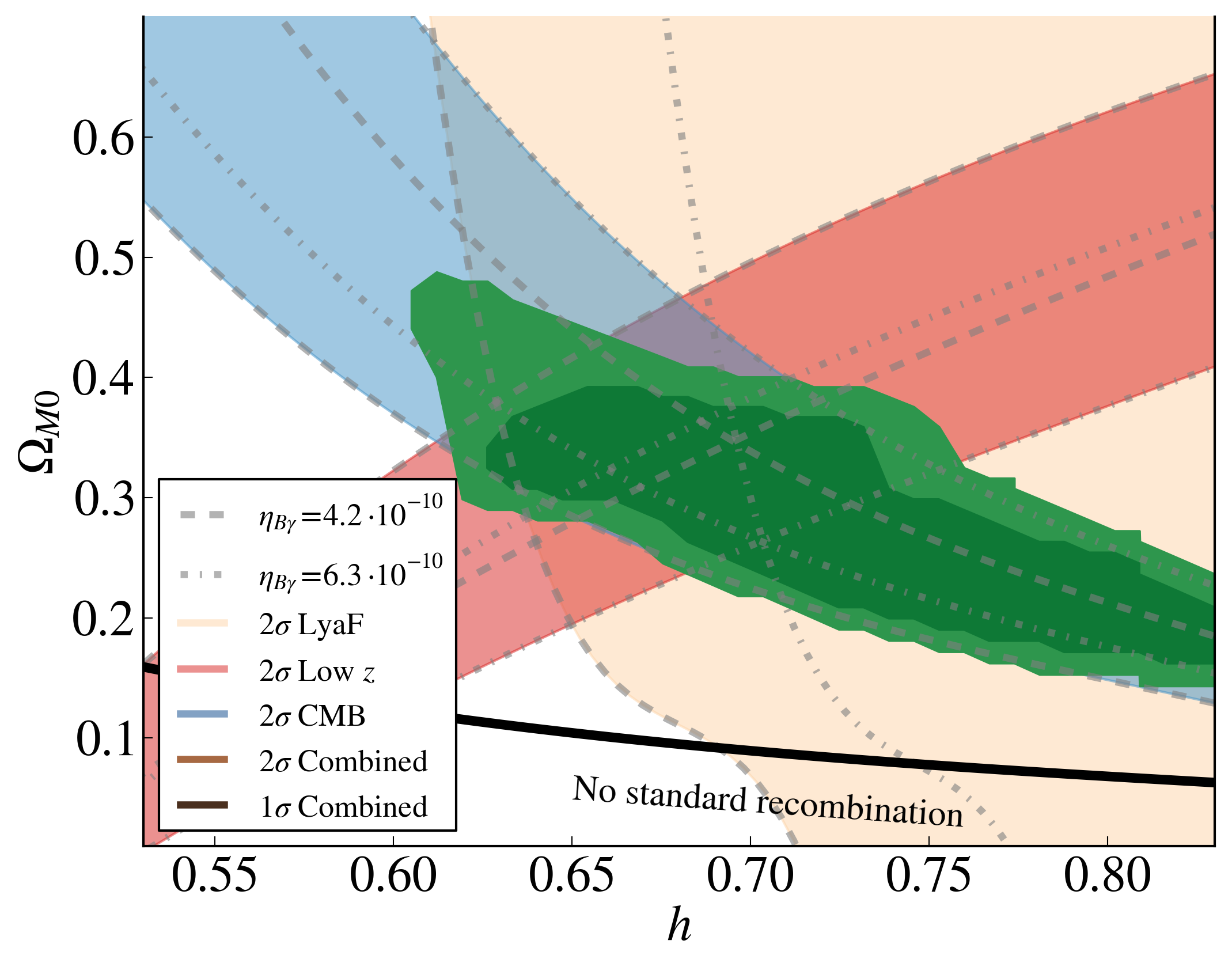}}
\caption{Cosmological parameter constraint priors from on the \ts\ model
(left panel) and the spatially flat \LCDM\ model (right panel). Independent
2$\si$ constraints are determined for: (i) the angular scale of sound
horizon in the CMB (contours from top left to bottom right); (ii) the fit of
the angular BAO scale from BOSS galaxies at low redshift (contours from
bottom left to top right); (iii) the angular BAO scale from one measurement
of the Lyman-$\alpha$ forest at $z=2.34$ (wide contours). A range of possible
baryon--to--photon ratios are allowed, with the extremes indicated. The
joint confidence region is determined by applying the CMB constraint and
allowing one or other BAO constraint.
\label{fig:prior}}
\end{figure*}
\subsection{Baryon Acoustic Oscillation constraints}
Determinations of the BAO scale from galaxy clustering at low redshifts
and Lyman alpha forest statistics at $z=2.34$ provide complementary
constraints on the expansion history. In previous
work \citep{dnw,lnw,sw} we simply demanded that the \ts\
effective comoving BAO scale match a single estimate determined from the FLRW
cosmology to within $\pm2$\%, $\pm4$\% or $\pm6$\%, which was a crude
method but the best available given the earlier precision of measurements.
The number and precision of measurements has now improved.

For the present investigation, we have considered estimates of the BAO scale at
different redshifts \citep{LymanA,BOSS1,BOSS2} using the best available data
from the BOSS survey. Unfortunately the standard FLRW cosmology plays an
implicit role in the data reduction, and limits the extent to which bounds
can be placed on non-FLRW models. The systematic issues can be most directly
understood by noting that the BAO scale is determined separately in the
angular and radial directions, by converting angular separations and redshift
separations in the galaxy--galaxy correlation function into the displacements
\beq \label{alphas}
\al_\perp=\frac{\left[\dA(z)/r_d\right]}{\left[\dA(z)/r_d\right]\ns{fid}}
\qquad\hbox{and}\qquad
\al_\parallel=\frac{\left[\dH(z)/r_d\right]}{\left[\dH(z)/r_d\right]\ns{fid}}
\eeq
where $r_d$ is the present comoving scale of the sound horizon at the
baryon drag epoch, $d_H(z)\equiv c/H(z)$, and the subscript ``fid'' refers
to quantities computed in a fiducial FLRW model that is used to convert
the raw angular and redshift displacements into 3-dimensional comoving
space. (Here we neglect the effect of redshift--space distortions which are
also often modelled with $N$-body Newtonian simulations based on the \LCDM\
model.)

The conversion to 3-dimensional comoving space can be
problematic for a non-FLRW model. While use of purely angular results
should pose no problems for the \ts\ cosmology, the conversion of redshift
increments to a radial comoving distance involves different assumptions about
spatial curvature in the FLRW and \ts\ models. One could in principle use
the values determined by a fiducial \LCDM\ model \citep{LymanA,BOSS1,BOSS2}
to recompute the radial comoving distance except for an additional problem:
in particular redshift ranges the relative Alcock--Paczy\'nski factor
$\left[\al_\perp/\al_\parallel\right]\Ns{\LA CDM}/
\left[\al_\perp/\al_\parallel\right]\Ns{TS}=\left[H(z)\dA(z)\right]\Ns{\LA CDM}/
\left[H(z)\dA(z)\right]\Ns{TS}$
between a fiducial \LCDM\ model and the \ts\ model varies over the redshift
slices $\DE z\goesas0.2$ used in the BOSS survey \citep{BOSS2} by an amount
similar in magnitude to the uncertainty. Consequently, to have
any confidence in radial measurements, one really needs to recompute the
radial BAO scale from the raw data assuming a fiducial \ts\ model. That
is beyond the scope of the present paper.

For the present analysis we will consequently restrict constraints on
the BAO scale to 2$\si$ bounds obtained from the angular estimates of
BOSS data \citep{BOSS2} at low redshifts $0.38\lsim z\lsim0.61$ and in
the Lyman-$\al$ forest \citep{LymanA} at $z=2.34$. In the former case,
the radial and angular measurements are actually also somewhat correlated.
Consequently, and also in view of the fact that the measurements at low
and high redshifts are in tension in the \LCDM\ model, we will take
bounds that result from the union of the constraints at low and high
redshifts, rather than their intersection. In practice, the bounds
are mostly set by the Lyman-$\al$ measurement since it has a much larger
uncertainty.
\subsection{Joint constraints}

The joint 2$\si$ confidence regions from applying the CMB constraint
and either the low-$z$ or $z=2.34$ BAO constraint is shown in
Fig.~\ref{fig:prior}. Since a range of possible baryon--to--photon ratios are
admitted, with no information from the relative heights of the acoustic peaks
used in either model, the width of the allowed regions is larger
than in conventional analyses for \LCDM.

For \ts\ the confidence regions are $\fvn\in(0.588,0.765)$ at 1$\si$,
$\fvn\in(0.500,0.799)$ at 2$\si$, $\fvn\in(0.378,0.826)$ at 3$\si$.
The corresponding effective dressed $\OmMn=\frn12(1-\fvn)(2+\fvn)$ is
$\OmMn\in(0.325,0.534)$ at 1$\si$, $\OmMn\in(0.281,0.625)$ at 2$\si$,
and $\OmMn\in(0.245,0.740)$ at 3$\si$.
For spatially flat \LCDM\ the corresponding confidence regions are
$\OmMn\in(0.162,0.392)$ at 1$\si$, $\OmMn\in(0.143,0.487)$ at 2$\si$,
and $\OmMn\in(0.124,0.665)$ at 3$\si$.
We adopt the 2$\si$ bounds as priors.


\begin{thebibliography}{99}
\expandafter\ifx\csname natexlab\endcsname\relax\def\natexlab#1{#1}\fi
\bibitem[Ade \etal(2016)]{planck}
Ade~P.~A.~R.~\etal, 
\art{Planck 2015 results. XIII. Cosmological parameters}
2016, \AaA{594}, A13 

\bibitem[Aghamousa \etal(2015)]{asas}
Aghamousa~A., Shafieloo~A., Arjunwadkar~M., Souradeep~T.,
\art{Unveiling acoustic physics of the CMB using nonparametric estimation of the temperature angular power spectrum for Planck}
2015, \JCAP{02}007. 

\bibitem[Akaike(1974)]{aic}
Akaike~H.,
\art{A new look at the statistical model identification}
1974, {IEEE Trans.\ Automat.\ Contr.}, {\bo19}, 716

\bibitem[Alam \etal(2017)]{BOSS2}
Alam~S.~\etal, 
\art{The clustering of galaxies in the completed SDSS-III Baryon Oscillation Spectroscopic Survey: cosmological analysis of the DR12 galaxy sample}
2017, \MNRAS{470}, 2617 

\bibitem[Aubourg \etal(2015)]{BOSS1}
Aubourg, \'E.~\etal, 
\art{Cosmological implications of baryon acoustic oscillation measurements}
2015, \PR D{92}, 123516 

\bibitem[Bennett \etal(2003)]{wmap}
Bennett~C.~L.~\etal,
\art{First Year Wilkinson Microwave Anisotropy Probe (WMAP) observations: Preliminary maps and basic result}
2003, \ApJs{148}, 1 

\bibitem[Betoule \etal(2014)]{jla}
Betoule~M.~\etal,
\art{Improved cosmological constraints from a joint analysis of the SDSS-II and SNLS supernova samples}
2014, \AaA{568}, A22 

\bibitem[Bolejko \etal(2016)Bolejko, Nazer \& Wiltshire]{bnw}
Bolejko~K., Nazer~M.~A., Wiltshire~D.~L.,
\art{Differential cosmic expansion and the Hubble flow anisotropy}
2016, \JCAP{06}035. 

\bibitem[Buchert(2000)]{be1}
Buchert~T.,
\art{On average properties of inhomogeneous fluids in general relativity. I: Dust cosmologies}
2000, \GRG{32}, 105 

\bibitem[Buchert(2001)]{be2}
Buchert~T.,
\art{On average properties of inhomogeneous fluids in general relativity. II: Perfect fluid cosmologies}
2001, \GRG{33}, 1381 

\bibitem[Buchert \& Carfora(2002)]{bc1}
Buchert~T., Carfora~M.,
\art{Regional averaging and scaling in relativistic cosmology}
2002, \CQG{19}, 6109 

\bibitem[Buchert \& Carfora(2003)]{bc2}
Buchert~T., Carfora~M.,
\art{Cosmological parameters are dressed}
2003, \PRL{90}, 031101 

\bibitem[Buchert \etal(2016)]{lcdm2}
Buchert~T., Coley~A.~A., Kleinert~H., Roukema~B.~F., Wiltshire~D.~L.,
\art{Observational Challenges for the Standard FLRW Model}
2016, {Int.\ J.\ Mod.\ Phys. D}, {\bo25}, 1630007 

\bibitem[Buchert \& R\"as\"anen(2012)]{BR}
Buchert~T., R\"as\"anen~S.,
\art{Backreaction in late-time cosmology}
2012, Ann.\ Rev.\ Nucl.\ Part.\ Sci.\ {\bo62}, 57 

\bibitem[Buchner \etal(2014)]{pymultinest}
Buchner~J.~\etal,
\art{X-ray spectral modelling of the AGN obscuring region in the CDFS: Bayesian model selection and catalogue}
2014, \AaA{564}, A125 

\bibitem[Bull \etal(2016)]{lcdm1}
Bull~P.~\etal,
\art{Beyond $\Lambda$CDM: Problems, solutions, and the road ahead}
2016, {Phys.\ Dark Univ.}, {\bo12}, 56 

\bibitem[Calcino \& Davis(2017)]{cd}
Calcino~J., Davis~T.,
\art{The need for accurate redshifts in supernova cosmology}
2017, \JCAP{01}038. 

\bibitem[Clarkson \etal(2008)Clarkson, Bassett \& Lu]{cbl}
Clarkson~C., Bassett~B., Lu~T.~H.~C.,
\art{A general test of the Copernican Principle}
2008, \PRL{101}, 011301 

\bibitem[Conley \etal(2007)]{c07}
Conley~A., Carlberg~R.~G., Guy~J., Howell~D.~A., Jha~S., Riess~A.~G., Sullivan~M.,
\art{Is there evidence for a Hubble bubble? The nature of type Ia supernova colors and dust in external galaxies}
2007, \ApJ{664}, L13

\bibitem[Conley \etal(2011)]{c11}
Conley~A.~\etal, 
\art{Supernova constraints and systematic uncertainties from the first 3 years of the Supernova Legacy Survey}
2011, \ApJs{192}, 1 

\bibitem[Cyburt \etal(2008)Cyburt, Fields \& Olive]{lithium}
Cyburt~R.~H., Fields~B.~D., Olive~K.~A.,
\art{An update on the big bang nucleosynthesis prediction for ${}
^7$Li: The problem worsens}
2008, \JCAP{11}012 

\bibitem[Delubac \etal(2015)]{LymanA}
Delubac~T.~\etal, 
\art{Baryon acoustic oscillations in the Ly$\alpha$ forest of BOSS DR11 quasars}
2015, \AaA{574}, A59 

\bibitem[Duley \etal(2013)Duley, Nazer \& Wiltshire]{dnw}
Duley~J.~A.~G., Nazer~M.~A., Wiltshire~D.~L.,
\art{Timescape cosmology with radiation fluid}
2013, \CQG{30}, 175006 

\bibitem[Ellis(1984)]{fit1}
Ellis~G.~F.~R.,
\art{Relativistic cosmology: Its nature, aims and problems}
In Bertotti, B., de Felice, F., Pascolini, A. (eds). {\em General Relativity and Gravitation}, (Reidel, Dordrecht, 1984) pp.~215--288.

\bibitem[Ellis \& Stoeger(1987)]{fit2}
Ellis~G.~F.~R., Stoeger~W.,
\art{The fitting problem in cosmology}
1987, \CQG{4}, 1697

\bibitem[Ellis \& Stoeger(2009)]{es09}
Ellis~G.~F.~R., Stoeger~W.,
\art{The evolution of our local cosmic domain: Effective causal limits}
2009, \MNRAS{398}, 1527 

\bibitem[Feroz \etal(2009)Feroz, Hobson \& Bridges]{multinest}
Feroz~F., Hobson~M.~P., Bridges~M.,
\art{MultiNest: an efficient and robust Bayesian inference tool for cosmology and particle physics}
2009, \MNRAS{398}, 1601 

\bibitem[Fixsen \etal(1996)]{f96}
Fixsen~D.~J., Cheng~E.~S., Gales~J.~M., Mather~J.~C., Shafer~R.~A., Wright~E.~L.,
\art{The cosmic microwave background spectrum from the full COBE/FIRAS data set}
1996, \ApJ{473}, 576

\bibitem[Guy \etal(2005)]{Guy05}
Guy~J., Astier~P., Nobili~S., Regnault~N., Pain~R.,
\art{SALT: a spectral adaptive light curve template for type Ia supernovae}
2005, \AaA{443}, 781 

\bibitem[Guy \etal(2007)]{Guy07}
Guy~J.~\etal,
\art{SALT2: using distant supernovae to improve the use of Type Ia supernovae as distance indicators}
2007, \AaA{466}, 11 

\bibitem[Haridasu \etal(2017)]{hldv}
Haridasu~B.~S., Lukovi\'c~V.~V., D'Agostino~R., Vittorio~N.,
\art{Strong evidence for an accelerating universe}
2017, \AaA{600}, L1 

\bibitem[Hicken \etal(2009)]{h09}
Hicken~M.~\etal, 
\art{Improved dark energy constraints from $\simeq$100 new CfA supernova type Ia light curves}
2009, \ApJ{700}, 1097

\bibitem[Hogg \etal(2005)]{h05}
Hogg~D.~W., Eisenstein~D.~J., Blanton~M.~R., Bahcall~N.~A., Brinkmann~J., Gunn~J.~E., Schneider~D.~P.,
\art{Cosmic homogeneity demonstrated with luminous red galaxies}
2005, \ApJ{624}, 54 

\bibitem[Hudson \etal(2004)]{hslb4}
Hudson~M.~J., Smith~R.~J., Lucey~J.~R., Branchini~E.,
\art{Streaming motions of galaxy clusters within 12000 km s$^{-1}$. 5. The peculiar velocity field}
2004, \MNRAS{352}, 61 

\bibitem[Jha \etal(2007)Jha, Riess \& Kirshner]{JRK}
Jha~S., Riess~A.~G., Kirshner~R.~P.,
\art{Improved distances to type Ia supernovae with Multicolor Light Curve Shapes: MLCS2k2}
2007, \ApJ{659}, 122 

\bibitem[Kass \& Raftery(1995)]{kr}
Kass~R.~E., Raftery~A.~E.,
\art{Bayes factors}
1995, {J.\ Am.\ Statist.\ Assoc.}, {\bo90}, 773

\bibitem[Larena \etal(2009)]{l09}
Larena~J., Alimi~J.-M., Buchert~T., Kunz~M., Corasaniti~P.~S.,
\art{Testing backreaction effects with observations}
2009, \PR D{79}, 083011 

\bibitem[Lavinto \etal(2013)Lavinto, R\"as\"anen \& Szybka]{lrs}
Lavinto~M., R\"as\"anen~S., Szybka~S.~J.,
\art{Average expansion rate and light propagation in a cosmological Tardis spacetime}
2013, \JCAP{12}051 

\bibitem[Leith \etal(2008)Leith, Ng \& Wiltshire]{lnw}
Leith~B.~M., Ng~S.~C.~C., Wiltshire~D.~L.,
\art{Gravitational energy as dark energy: Concordance of cosmological tests}
2008, \ApJ{672}, L91 

\bibitem[Lewis \etal(2016)Lewis, Barnes \& Kaushik]{lbk}
Lewis~G.~F., Barnes~L.~A., Kaushik~R.,
\art{Primordial nucleosynthesis in the $R_h = ct$ cosmology: Pouring cold water on the simmering Universe}
2016, \MNRAS{460}, 291 

\bibitem[March \etal(2011)]{mtbsv}
March~M.~C., Trotta~R., Berkes~P., Starkman~G.~D., Vaudrevange~P.~M.,
\art{Improved constraints on cosmological parameters from SNIa data}
2011, \MNRAS{418}, 2308 

\bibitem[McKay \& Wiltshire(2016)]{mw}
McKay~J.~H., Wiltshire~D.~L.,
\art{Defining the frame of minimum nonlinear Hubble expansion variation}
2016, \MNRAS{457}, 3285 (2016); err.\ {\bo463}, 3113 

\bibitem[Mosher \etal(2014)]{m14}
Mosher~J. \etal,
\art{Cosmological parameter uncertainties from SALT-II Type Ia supernova light curve models}
2014, \ApJ{793}, 16 

\bibitem[Nazer \& Wiltshire(2015)]{nw}
Nazer~M.~A., Wiltshire~D.~L.,
\art{Cosmic microwave background anisotropies in the timescape cosmology}
2015, \PR D{91}, 063519 

\bibitem[Nielsen \etal(2016)Nielsen, Guffanti \& Sarkar]{ngs}
Nielsen~J.~T., Guffanti~A., Sarkar~S.,
\art{Marginal evidence for cosmic acceleration from Type Ia supernovae}
2016, {Sci.\ Rep.}, {\bo6}, 35596 

\bibitem[Olive \etal(2000)Olive, Steigman \& Walker]{bbn1}
Olive~K.~A., Steigman~G., Walker~T.~P.,
\art{Primordial nucleosynthesis: Theory and observations}
2000, {Phys.\ Rept.}, {\bo333}, 389 

\bibitem[Perlmutter \etal(1999)]{P98}
Perlmutter~S.~\etal, 
\art{Measurements of Omega and Lambda from 42 high-redshift supernovae}
1999, \ApJ{517}, 565 

\bibitem[R\"as\"anen \etal(2015)R\"as\"anen, Bolejko \& Finoguenov]{rbf}
R\"as\"anen~S., Bolejko~K., Finoguenov~A.,
\art{New test of the Friedmann-Lema\^{\i}
tre-Robertson-Walker metric using the distance sum rule}
2015, \PRL{115}, 101301 

\bibitem[Riess \etal(1998)]{R98}
Riess~A.~G.~\etal, 
\art{Observational evidence from supernovae for an accelerating universe and a cosmological constant}
1998, \AJ{116}, 1009 

\bibitem[Riess \etal(2007)]{R07}
Riess~A.~G.~\etal,
\art{New Hubble Space Telescope discoveries of Type Ia supernovae at $z > 1$: Narrowing constraints on the early behavior of dark energy}
2007, \ApJ{659}, 98 

\bibitem[Rubin \& Hayden(2016)]{rh}
Rubin~D., Hayden~B.,
\art{Is the expansion of the universe accelerating? All signs point to yes}
2016, \ApJ{833}, L30 

\bibitem[Sapone \etal(2014)Sapone, Majerotto \& Nesseris]{smn}
Sapone~D., Majerotto~E., Nesseris~S.,
\art{Curvature versus distances: Testing the FLRW cosmology}
2014, \PR D{90}, 023012 

\bibitem[Schwarz(1978)]{bic}
Schwarz~G.,
\art{Estimating the dimension of a model}
1978, {Ann.\ Statist.}, {\bo6}, 461

\bibitem[Scrimgeour \etal(2012)]{sdb12}
Scrimgeour~M.~\etal, 
\art{The WiggleZ Dark Energy Survey: the transition to large-scale cosmic homogeneity}
2012, \MNRAS{425}, 116 

\bibitem[Shariff \etal(2016)]{bahamas}
Shariff~H., Jiao~X., Trotta~R., van Dyk~D.~A.,
\art{BAHAMAS: New analysis of type Ia supernovae reveals inconsistencies with standard cosmology}
2016, \ApJ{827}, 1 

\bibitem[Skilling(2004)]{SkillingNS}
Skilling~J.,
\art{Nested sampling}
2004, {AIP Conf.\ Proc.}, {\bo735}, 395

\bibitem[Smale(2011)]{grb}
Smale~P.~R.,
\art{Gamma-ray burst distances and the timescape cosmology}
2011, \MNRAS{418}, 2779 

\bibitem[Smale \& Wiltshire(2011)]{sw}
Smale~P.~R., Wiltshire~D.~L.,
\art{Supernova tests of the timescape cosmology}
2011, \MNRAS{413}, 367 

\bibitem[Spiegelhalter \etal(2002)]{spiegelhalter2002}
Spiegelhalter~D.~J., Best~N.~G., Carlin~B.~P., van der Linde~A.,
\art{Bayesian measures of model complexity and fit}
2002, {J.\ Royal.\ Stat.\ Soc.\ B}, {\bo64}, 583

\bibitem[Trotta(2007)]{t07}
Trotta~R.,
\art{Applications of Bayesian model selection to cosmological parameters}
2007, \MNRAS{378}, 72

\bibitem[Tutusaus \etal(2017)]{tldb}
Tutusaus~I., Lamine~B., Dupays,~A., Blanchard~A.,
\art{Is cosmic acceleration proven by local cosmological probes?}
2017, \AaA{602}, A73 

\bibitem[Tytler \etal(2000)]{bbn2}
Tytler~D., O'Meara~J.~M., Suzuki~N., Lubin~D.,
\art{Review of big bang nucleosynthesis and primordial abundances}
2000, {Phys.\ Scripta}, {\bo T85}, 12 

\bibitem[van de Weygaert \etal(2016)]{web}
van de Weygaert~R., Shandarin~S., Saar~E., Einasto~J., eds.,
\art{The Zeldovich Universe: Genesis and growth of the cosmic web}
{Proc.\ IAU Symposium}, {\bo308}, (Cambridge Univ.\ Press, 2016)

\bibitem[Visser(2004)]{Vis04}
Visser~M.,
\art{Jerk and the cosmological equation of state}
2004, \CQG{21}, 2603 

\bibitem[Wiltshire(2007\natexlab{a})]{clocks}
Wiltshire~D.~L.,
\art{Cosmic clocks, cosmic variance and cosmic averages}
2007\natexlab{a}, New J.\ Phys.\ {\bo9}, 377 

\bibitem[Wiltshire(2007\natexlab{b})]{sol}
Wiltshire~D.~L.,
\art{Exact solution to the averaging problem in cosmology}
2007\natexlab{b}, \PRL{99}, 251101 

\bibitem[Wiltshire(2008)]{cep}
Wiltshire~D.~L.,
\art{Cosmological principle of equivalence and the weak--field limit}
2008, \PR D{78}, 084032 

\bibitem[Wiltshire(2009)]{obs}
Wiltshire~D.~L.,
\art{Average observational quantities in the timescape cosmology}
2009, \PR D{80}, 123512 

\bibitem[Wiltshire(2011)]{dust}
Wiltshire~D.~L.,
\art{What is dust? -- Physical foundations of the averaging problem in cosmology}
2011, \CQG{28}, 164006 

\bibitem[Wiltshire(2014)]{bscg}
Wiltshire~D.~L.,
\art{Cosmic structure, averaging and dark energy}
In Novello, M., Perez Bergliaffa, S.E., eds. {\em Cosmology and Gravitation: XVth Brazilian School of Cosmology and Gravitation}, (Cambridge Scientific Publishers, 2014), pp 203-244; [arXiv:1311.3787].

\bibitem[Wiltshire \etal(2013)]{rest}
Wiltshire~D.~L., Smale~P.~R., Mattsson~T., Watkins~R.,
\art{Hubble flow variance and the cosmic rest frame}
2013, \PR D{88}, 083529 
\end{thebibliography}
\end{document}